\documentclass
[10pt,conference]{IEEEtran}

\usepackage[normalem]{ulem}
\usepackage{mathptmx} %
\usepackage{fancyhdr}
\usepackage[normalem]{ulem}
\usepackage{upgreek} %
\usepackage{graphicx}
\usepackage[dvipsnames]{xcolor}
\usepackage{color, soul}
\usepackage[ruled, linesnumbered, vlined]{algorithm2e}
\usepackage{xspace}
\usepackage{multirow}
\usepackage{array}
\usepackage{comment}
\usepackage{listings}
\usepackage{microtype}
\usepackage{graphicx}
\usepackage{booktabs} %
\usepackage[normalem]{ulem}
\usepackage{lipsum} %
\usepackage{comment} %
\usepackage{xcolor}
\usepackage{xspace}
\usepackage{microtype}
\usepackage{graphicx}
\usepackage{scalerel}
\usepackage{booktabs}
\usepackage{tabularx}
\usepackage{enumitem}
\usepackage{balance} 
\usepackage{tikz} %
\usepackage{caption}
\usepackage{adjustbox}
\usepackage{subfig}
\usepackage{url} 
\usepackage{dblfloatfix}
\usepackage{enumitem}
\usepackage{subfig}
\usepackage[bookmarks=true,letterpaper=true,colorlinks,linkcolor=black,citecolor=blue,urlcolor=black]{hyperref}
\usepackage[capitalize,noabbrev,nameinlink]{cleveref}

\lstdefinestyle{CStyle}{
    backgroundcolor=\color{white},   
    commentstyle=\color{mGreen},
    keywordstyle=\color{blue},
    numberstyle=\tiny\color{mGray},
    stringstyle=\color{mPurple},
    basicstyle=\scriptsize\ttfamily,
    breakatwhitespace=false,         
    breaklines=true,                 
    captionpos=b,                    
    keepspaces=true,                 
    numbers=left,                    
    numbersep=3pt,                  
    showspaces=false,                
    showstringspaces=false,
    showtabs=true,                  
    tabsize=1,
    language=C,
    float=tp
}
\crefformat{section}{#2§#1#3}
\crefformat{subsection}{#2\S#1#3}
\crefformat{subsubsection}{#2\S#1#3}
\Crefname{figure}{Fig.}{Figs.}
\Crefname{equation}{Eqn.}{Eqns.}
\Crefname{section}{\S}{\S}

\setstcolor{red}

\newcommand{\hpcayear}{2024}

\title{\TheName: A Comprehensive Cluster Design Methodology for Distributed Deep Learning Training}

\def\distrib{} %
\newcommand\hpcaauthors{Divya Kadiyala, Saeed Rashidi, Taekyung Heo, Abhimanyu Bambhaniya, Tushar Krishna, Alexandros Daglis}
\newcommand\hpcaaffiliation{Georgia Institute of Technology}

\newcommand{\astrasim}{\textsc{ASTRA-sim}\xspace}

\newcommand{\TheName}{{COMET}\xspace} 
\newcommand{\TheModel}{Transformer-1T\xspace}

\newcommand{\TextCircled}[1]{{\large{\textcircled{\normalsize \textbf{#1}}}}}

\newcommand{\revision}[1]{#1}

\author{
  \ifdefined\hpcacameraready
    \IEEEauthorblockN{\hpcaauthors{}}
      \IEEEauthorblockA{
        \hpcaaffiliation{} \\
        \hpcaemail{}
      }
  \fi
  \ifdefined\distrib
    \IEEEauthorblockN{\hpcaauthors{}}
      \IEEEauthorblockA{
        \hpcaaffiliation{} 
      }
  \else
    \IEEEauthorblockN{\normalsize{HPCA \hpcayear{} Submission
      \textbf{\#\hpcasubmissionnumber{}}} \\
      \IEEEauthorblockA{
        Confidential Draft \\
        Do NOT Distribute!!
        \vspace{-4mm}
      }
    }
  \fi 
}

\fancypagestyle{camerareadyfirstpage}{%
  \fancyhead{}
  
  \fancyhead[C]{
  \ifdefined\distrib
    \else
    \ifdefined\aeopen
    \parbox[][12mm][t]{13.5cm}{\hpcayear{} IEEE International Symposium on High-Performance Computer Architecture (HPCA)}    
    \else
      \ifdefined\aereviewed
      \parbox[][12mm][t]{13.5cm}{\hpcayear{} IEEE International Symposium on High-Performance Computer Architecture (HPCA)}
      \else
      \ifdefined\aereproduced
      \parbox[][12mm][t]{13.5cm}{\hpcayear{} IEEE International Symposium on High-Performance Computer Architecture (HPCA)}
      \else
      \parbox[][0mm][t]{13.5cm}{\hpcayear{} IEEE International Symposium on High-Performance Computer Architecture (HPCA)}
    \fi 
    \fi 
    \fi 
    \ifdefined\aeopen 
      \includegraphics[width=12mm,height=12mm]{ae-badges/open-research-objects.pdf}
    \fi 
    \ifdefined\aereviewed
      \includegraphics[width=12mm,height=12mm]{ae-badges/research-objects-reviewed.pdf}
    \fi 
    \ifdefined\aereproduced
      \includegraphics[width=12mm,height=12mm]{ae-badges/results-reproduced.pdf}
    \fi
    \fi
  }
  \ifdefined\distrib
  \else
    \fancyfoot[C]{}
  \fi
}
\begin{document}
\maketitle

\ifdefined\distrib
\thispagestyle{camerareadyfirstpage}
  \pagestyle{plain}
\else 
\ifdefined\hpcacameraready 
  \thispagestyle{camerareadyfirstpage}
  \pagestyle{empty}
\else
  \thispagestyle{plain}
  \pagestyle{plain}
\fi
\fi

\newcommand{\hpcaheight}{0mm}
\ifdefined\eaopen
\renewcommand{\hpcaheight}{12mm}
\fi

\begin{abstract}
Modern Deep Learning (DL) models have grown to sizes requiring massive clusters of specialized, high-end nodes to train.
Designing such clusters to maximize both performance and utilization---to amortize their steep cost---is a challenging task requiring careful balance of compute, memory, and network resources.
Moreover, a plethora of each model’s tuning knobs drastically affect the performance, with optimal values often depending on the underlying cluster’s characteristics, which necessitates a complex cluster-workload co-design process. 
To facilitate the design space exploration of such massive DL training clusters, we introduce \TheName, a holistic cluster design methodology and workflow to jointly study the impact of parallelization strategies and key cluster resource provisioning on the performance of distributed DL training. 
We develop a step-by-step process to establish a reusable and flexible methodology, and demonstrate its application with case studies of training large models on cluster configurations of variable compute, memory, and network resources. 
Our case studies demonstrate \TheName’s utility in identifying promising architectural optimization directions and guiding system designers in configuring key model and cluster parameters.
To illustrate, cluster configuration comparisons identify performance differences of up to $7.7\times$ and highlight performance optimization opportunities of up to  $1.4\times$ when employing memory expansion as an optimization technique.

\end{abstract}
\section{Introduction}
\label{sec:intro}

Modern Deep Learning (DL) workloads such as natural language processing \cite{smith:megatron-NLG, DBLP:wolf:NLP}, drug discovery \cite{Grechishnikova:drug-discovery, vamathevan:drug-discovery}, text-to-speech  conversion \cite{ amodei:deepspeech2, ren:fastspeech, DBLP:thoppilan:lamda}, and personal recommendation engines \cite{DBLP:maxim:DLRM, zhang:DLRM-Survey} are becoming increasingly pervasive and commercially important, forming the core components of many day-to-day applications and services deployed in datacenters. 
The rapid growth of these models %
has culminated in massive resource requirements (terabytes of memory, petaflops of compute) to train in a practical timeframe.
Training is therefore conducted in a distributed fashion over massive clusters of high-end specialized accelerator nodes, such as GPUs or TPUs, connected over high-bandwidth networks \cite{jouppi:tpuv4, narayanan:efficient}.
Given {the abundance of available compute (GPU, TPU, custom accelerator), network (Ethernet, InfiniBand, NVLink), and memory system (HBM, DRAM over DDR or CXL) technologies, as well as} the steep cost of such clusters, designing them for maximum performance and efficiency is a {challenge of high complexity and critical importance.}

Optimizing a cluster for distributed DL training requires keen understanding of key model characteristics, training strategies, and hardware components.
Maximizing efficiency not only requires carefully balancing the cluster's compute, memory, and network characteristics, but also adjusting the model's parallelization strategy to best suit those underlying hardware resources.
Therefore, a holistic, end-to-end methodology is needed to analyze and understand the impact of different system components and training strategies on each cluster's DL training performance and efficiency.

\begin{figure}[t]
    \centering
    \includegraphics[width=.8\columnwidth]{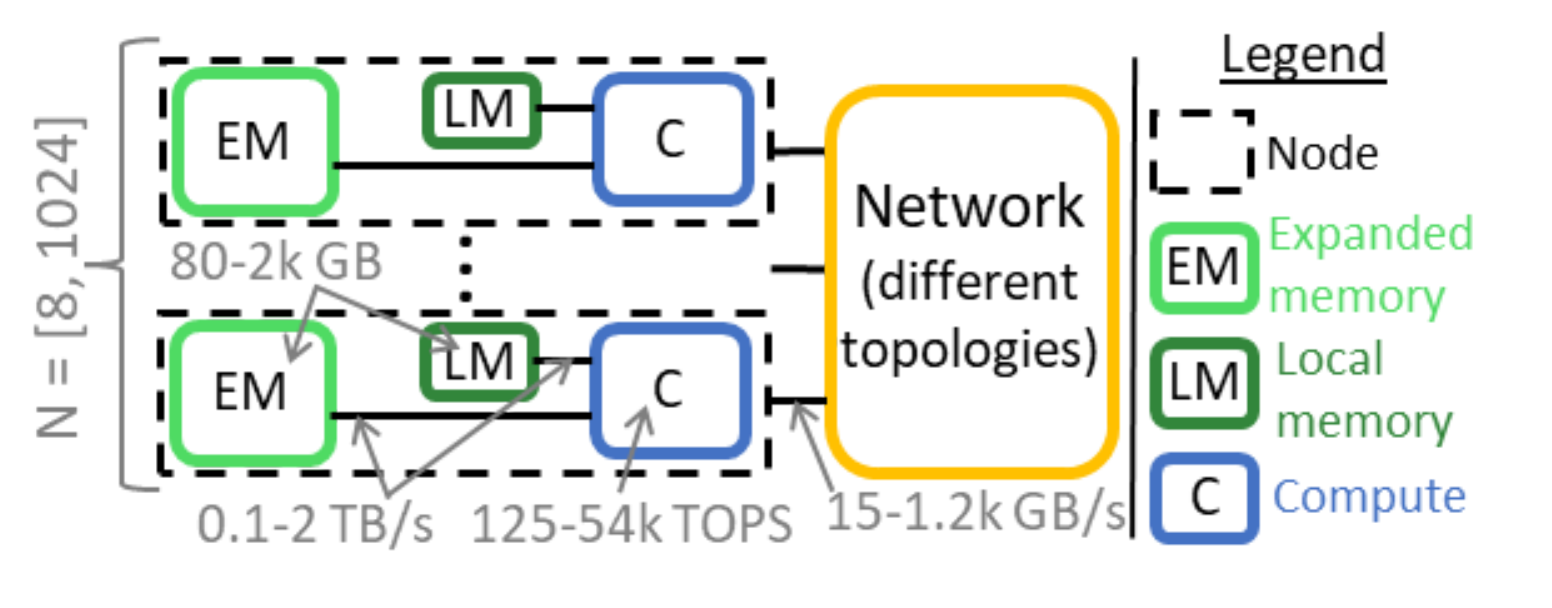}
     \vspace{-3mm}
    \caption{Tunable cluster component parameters in \TheName and value ranges evaluated in this paper.}
    \label{fig:eval-space}
    \vspace{-4mm}
\end{figure}

To address this need, we introduce the \TheName  methodology, 
which allows rapid joint exploration of cluster resource provisioning and training parallelization strategies. 
\TheName includes a detailed workflow to break down a model into its layers, analyze its training behavior as a function of the employed parallelization strategy, and assess the impact of a cluster's key resources %
on training performance and efficiency.
\TheName is a versatile tool that %
helps cluster designers determine the optimal resource provisioning balance for a set of target training algorithms. 
The methodology also enables researchers and technologists to study and quickly quantify the impact of emerging technologies or of any modification of a given component's characteristics (e.g., per-node compute density, memory capacity or bandwidth, network bandwidth or latency, etc.) on distributed DL training performance and efficiency.
For instance, {a timely question is whether upcoming CXL-enabled memory expansion can be leveraged to improve a cluster's DL training performance.
We use \TheName to determine the capacity and bandwidth characteristics required by such a new memory solution to be impactful.
To illustrate, \cref{fig:eval-space} summarizes the hardware components of the clusters modeled in \TheName, along with the range of values evaluated in this paper.
} 
Overall, by enabling rapid and holistic studies, \TheName informs cluster designers with a resource provisioning balance that maximizes training efficiency for a target set of DL models.

\noindent In summary, we make the following contributions: %
\vspace{-0.5mm}
\begin{itemize}[noitemsep, topsep=0pt,leftmargin=*]
    \item We construct the holistic \TheName methodology to enable rapid design space {\textit{co-exploration} of model training parallelization strategies and key cluster resource parameters, to assess their joint impact} on the performance of distributed DL training.
    \item We implement our methodology with a streamlined toolchain %
    and showcase its utility with case studies using different large Deep Learning Recommendation Models (DLRM) and language (Transformer) models.
    \item \TheName helps system architects rapidly quantify the impact of various current and future accelerators, network capabilities, and memory system technologies on cluster performance and efficiency. We particularly emphasize the potential of memory expansion techniques (e.g., CXL-attached memory) as an understudied yet promising cluster design knob.     
\end{itemize}

\noindent\textbf{Paper outline:}
\cref{sec:background_motivation} covers background and related work.
\cref{sec:method} details \TheName's methodology and \cref{sec:instantiation} its implementation.
\cref{sec:case-studies} demonstrates \TheName's utility with  use cases and \cref{sec:conclusion} concludes.

\section{Background and Related Work}
\label{sec:background_motivation}

\subsection{Distributed DL Parallelization Strategies}
\label{sec:background:par-strategies}

In recent years, we have witnessed tremendous growth in DL model sizes, with current largest models already trending in the 100s of billions or  trillions of parameters~\cite{fedus:Switch-Transformers, Lin:M6, smith:megatron-NLG}.
Training models of such size requires tremendous computational and memory resources, only attainable on massive clusters with hundreds of (typically GPU-based) computing nodes.
Given a cluster deployment, there is a range of parallelization strategies to choose for distributed DL training such as Data Parallelism (DP) \cite{DBLP:Li:Data-Parallelism}, Model Parallelism (MP) \cite{shazeer:model-parallelism, DBLP:shoeybi:megatronlm}, Pipeline Parallelism (PP) \cite{DBLP:huang:GPipe}, and more general parallelization strategies (e.g., expert parallelism \cite{fedus:Switch-Transformers}). 
\TheName currently focuses on analyzing the effects of the two foundational parallelization strategies of MP and DP\footnote{{Fully Sharded Data Parallelism (FSDP) \cite{pytorchfsdp2,metafsdp,pytorchfsdp1} is a popular training strategy logically similar to DP, but stages data trough the GPU's local memory and its adjacent CPU host's memory. \TheName captures such staging between local and expanded memory, hence representing FSDP-style training as well.}}, but its modular design allows future extension to model other strategies as well.

In MP training, the model is split across the nodes, hence each node holds a shard of the model, requiring frequent inter-node communication both during forward and backward propagation.
In contrast, in DP training, each node holds the entire model and the training batch is sharded across nodes. 
Inter-node communication is only required on backward propagation to reconcile weight updates across the multiple model copies.
As a result, MP training generally requires more frequent synchronous inter-node communication, while DP training results in less frequent but more voluminous communication, which is easier to overlap with computations.

\subsection{Challenges in Distributed DL Training}
\label{sec:memory-role}
The performance and efficiency of a cluster used in distributed DL training is dictated by a range of key parameters: per-node computational capability, memory capacity and bandwidth, and the inter-node network bandwidth.
In addition to raw hardware capabilities, a training task's performance is affected by the training task's structure (i.e., the chosen parallelization strategy), as that affects the resulting workload characteristics and, consequently, how each cluster resource is being stressed.
Limiting our scope to the two foundational MP and DP parallelization strategies, the chosen MP/DP balance for a given training task drastically affects the workload's characteristics, bearing significant implications in the resulting computation/communication balance, per-node memory capacity requirements, and, ultimately, the distributed training's performance and efficiency.
Therefore, optimizing training performance requires holistic co-design of the parallelization strategy alongside the cluster's memory, compute, and network resources.
\revision{The \TheName methodology enables composite studies where cluster resource provisioning strategies and parallelization strategies for a training task are jointly varied.}

\subsection{Prior Work on Distributed DL Training}

\noindent\textbf{DL training accelerator design.} A vast body of prior work focuses on optimizing the individual accelerator node (GPU or custom accelerator) \cite{chen:diannao, chen:eyeriss, jouppi:TPUV3,  kwon:maeri, parashar:timeloop, qin:sigma, samajdar:scalesim}.
However, node design in isolation does not capture cluster-scale effects during distributed training and can lead to resource imbalance and cluster under-utilization. %
Maximizing cluster-wide performance and utilization requires a holistic design approach that jointly considers the impact of compute, network, memory, and workload parallelization strategy.

\noindent\textbf{Cluster-scale DL training performance analysis.}
Closer to our work, some prior work characterizes the performance of large-scale distributed training on current clusters.
Jain et al.  evaluate the performance of training ResNet and Inception-based models on different CPU and GPU architectures, as a function of  batch size, node count, and threads per node~\cite{jain:DNN-perf-characterization}. 
Ren et al.  analyze the distributed training performance of NLP and computer vision workloads leading-edge systems, providing insights into the collective communication kernels and the impact of node count scaling on overall training throughput~\cite{ren:PMBS}.
Jeon et al.  focus on multi-tenant GPU clusters hosting co-located DNN training workloads, study how locality-aware scheduling affects performance and utilization, and propose guidelines for improved cluster schedulers \cite{jeon:Philly}. %

\noindent\textbf{Cluster communication performance optimizations.} Another family of work focuses on cluster communication performance, which is often a major performance determinant.
Dong et al. propose an algorithm/system co-design methodology to improve training scalability, by alleviating network congestion with a congestion-less server architecture that uses novel communication collective algorithms and network topology~\cite{dong:EFLOPS}.
Jiang et al.  accelerate DNN training 
with a unified communication framework that dynamically adapts reduction collectives and parameter server tasks to best utilize CPU and bandwidth resources, and overlap communication latency~\cite{yimin:byteps}.
Shah et al. and Sun et al. propose communication abstraction and improvised communication algorithms~\cite{DBLP:shah:TACCL, sun:gradientFlow}.

\noindent\textbf{Memory system impact on training.} Memory system design and optimizations play a crucial role in the overall performance and efficiency of a distributed training cluster, as well as the cluster size required to train a given large model. Prior works such as Checkmate \cite{jain:checkmate}, ZeRO-DP \cite{rajbhandari:zeroDP}, ZeRO-Offload \cite{ren:zeRO-Offload}, and ZeRO-Infinity \cite{rajbhandari:zero-infinity}  focus on reducing the per-node memory footprint required to train large DL models.%

\noindent\textbf{Training strategy auto-tuning frameworks.}
Prior works on auto-tuning frameworks (Alpa \cite{zheng:2022}, AutoDDL~\cite{chen:autoddl}, Rhino \cite{zhang:rhino}, FlexFlow \cite{jia:flexflow}) focus on identifying the optimal parallelization strategy using a combination of data, operator and communication patterns.
These frameworks perform an exhaustive design space search to identify the optimal parallelization strategy for a \textit{given} cluster.
While auto-tuning frameworks predict the optimal parallelization strategy for a DL model training task on a \textit{specific} cluster, \TheName develops a generic methodology to jointly evaluate the performance of a training strategy on an \textit{arbitrary} cluster at an earlier design stage: when a cluster’s architecture is {still malleable, i.e.,  going under design considerations}. Our goal is \textit{not} to identify the best parallelization strategy for training a given DL model on an existing cluster, but rather to enable rapid design space exploration and evaluation of impact of various model parallelization strategies along with key architectural cluster design parameters. %
\revision{\TheName currently does not automate the workload and cluster design space exploration. A future extension could add a frontend layer in the toolchain that automates generation of workload input files (representing different parallelization strategies) and cluster configurations (i.e., steps \TextCircled{2} and \TextCircled{5} in \cref{fig:COMET-instance}, described in \cref{sec:instantiation}).}

\smallskip\noindent
All aforementioned categories of prior work provide valuable insights to understand the performance characteristics of individual components of a large training cluster.
However, a cluster design approach that \textit{jointly} considers the training parallelization strategy along with the cluster's key architectural parameters is essential to not only maximize performance and efficiency, but also identify the true impact any individual or combination of future changes will have on the metrics of interest. %
We, therefore, argue for a holistic cluster design methodology encapsulating: (i) the target model to be trained with the range of possible parallelization strategies, each node's (ii) compute and (iii) memory subsystem, and (iv) the cluster's network\footnote{\revision{In this paper, the term ``node''  refers to one compute unit (e.g., a GPU, a CPU, a TPU, etc.).}}.
Understanding these tradeoffs at scale is challenging, and we are not aware of a \textit{publicly available unified framework }that allows researchers to perform such broad design space exploration quickly. 
{\textit{\TheName is a methodology and toolchain that aims to fill that gap, facilitating rapid iterations of algorithm/hardware co-design for large-scale DL training.}}

\section{The \TheName Methodology}
\label{sec:method}

\begin{figure}[t]
    \centering
    \includegraphics[width=\columnwidth]{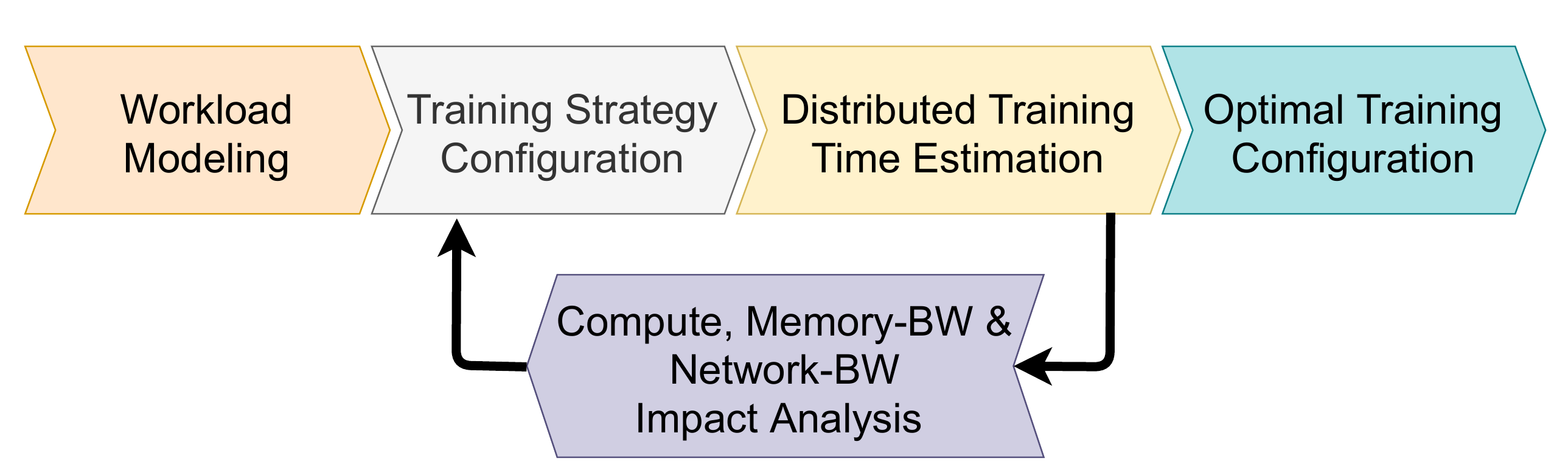}
    \caption{\TheName methodology overview.}
    \label{fig:method-overview}
    \vspace{-5mm}
\end{figure}

We design the \TheName methodology to holistically evaluate a large-scale model's training strategy on an arbitrary cluster and assess the best combination of training strategy and cluster resource balance to maximize training performance and/or cost efficiency.
\cref{fig:method-overview} shows a high-level overview of our methodology and this section details each of its steps.

\subsection{Workload Modeling}
\label{sec:method:workload-modeling}

The first step involves decomposing the model of interest into its layers, along with the number of operations and data movement requirements in each layer.
We express each layer as a general matrix-matrix multiplication (GEMM) between input activations ($M\times K$) and weights ($K\times N$), producing an output matrix ($M\times N$). %
After decomposing the model into layers, we compute the number of parameters for the operand matrices of each layer.
The total size of a model, in terms of number of parameters, is given by the sum of the weight matrices' (i.e., $K \times N$) elements \cite{vaswani:attention}. 
{Layers that cannot be encoded as GEMMs (e.g., embedding-lookups) are represented by their input/output operand sizes, total number of operations and data moved between memory and the compute unit.}

\begin{figure}[t]
    \centering    
    \includegraphics[width=\columnwidth]{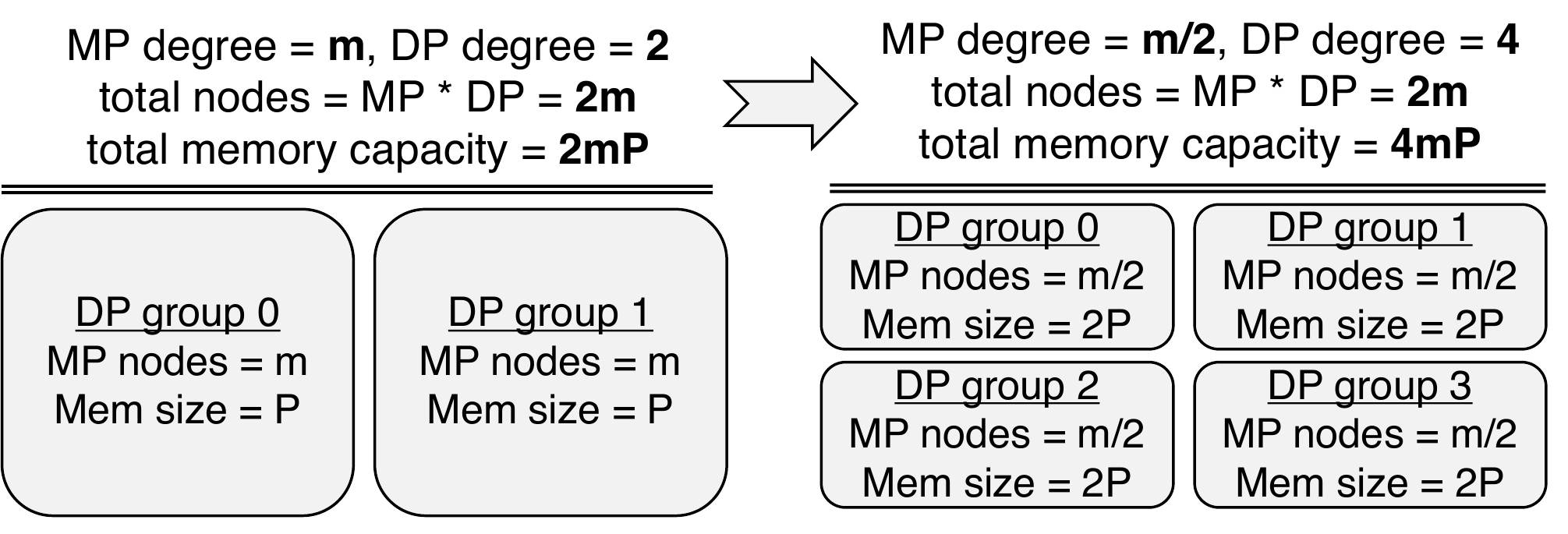}
    \caption{Variation of per-node memory capacity requirements as a function of MP and DP degrees in a fixed-size cluster. }
    \label{fig:isonode-mem-variation}
     \vspace{-3mm}
\end{figure}

\subsection{Training Strategy Configuration}
\label{sec:method:training-strategy}
In this step, we determine the parallelization strategy for the selected model based on the model type and per-node memory capacity.
Different strategies focus on optimizing training for different metrics such as throughput, compute utilization, inter-node communication, etc.
The current version of \TheName focuses on Data and Model parallelism (MP and DP---see \cref{sec:background:par-strategies}).
Given a target cluster size, we compute the per-node memory capacity requirement as a function of the selected degree of MP and DP in the cluster.
We compute the per-node memory footprint to hold all the data (model, input/output matrices) required for the distributed DL training task.
The model's memory footprint is dictated by model states and activations.
We compute the memory footprint required for each operand matrix based on its type (i.e., model weights or input/output activations), size of parameters, and the memory optimizations (e.g., ZeRO-DP, as explained later in Section \cref{sec:instantiation}).
For a cluster of size N, we start with the initial condition where all the nodes are within the MP dimension (i.e., $MP=N, DP=1$) and, collectively, hold a single copy of the entire model.
Then we sweep the (MP, DP) degree to the other extreme (i.e., $MP=1, DP=N$), considering all power-of-two combinations, with $MP \times DP = N$.

\revision{Doubling the DP dimension implies halving the MP dimension, so half the number of nodes must hold an entire copy of the model. Consequently, the per-node memory capacity requirement doubles.
To illustrate, \cref{fig:isonode-mem-variation} shows a cluster configured as two data-parallel node groups ($DP=2$) and each DP node group consisting of $m$ nodes performing $m$-way model parallelism ($MP=m$), for a total node count $N=2m$.
To contain a copy of the entire model of size C across each DP group's $m$ nodes, each node requires a minimum memory capacity $P=\frac{C}{m}$.
}
Moving from a $DP=2, MP=m$ to a $DP=4, MP=\frac{m}{2}$ configuration doubles the per-node memory capacity requirement to $2P$, as each DP group of  $\frac{m}{2}$ nodes must now hold the entire model $C$. 
As a result, the cluster's total memory capacity also doubles from $2mP$ to $4mP$.

The chosen (MP, DP) degree not only affects the memory requirement per node, but also results in different computation requirements and communication behavior.
Therefore, for each (MP, DP) combination, we also derive the volume of per-node computation and inter-node communication.

\subsection{Distributed Training Time Estimation}
\label{sec:method:training-time-estimation}
The computation and communication requirements per layer for each (MP, DP) combination are fed into a performance model that estimates the model's training time. 
The performance model estimates end-to-end training time as a function of the modeled cluster's per-node compute capability, memory capacity and bandwidth, network bandwidth and topology.

{A key design decision in \TheName is opting for generality and breadth rather than detailed modeling of individual components, as our methodology is intended to enable rapid and scalable exploration of a vast design space, rather than highly accurate performance estimations.
\TheName's goal is to allow gleaning \textit{performance trends} as cluster parameters are varied both jointly and separately.
Therefore, we chose to go with detailed analytical models for compute, memory, and network rather than be tied to any specific technology or component instance.
Performance estimation for DL training models lends itself well to analytical modeling (as opposed to cycle-level simulation), as their computation, memory access, and communication patterns exhibit regularity typically absent from general workloads~\cite{castello:performance, lu:distsim}. %
By sweeping generic characteristics like TOPS, bandwidth, latency, \TheName's users may easily create proxies for specific components or technologies of interest, such as GPUs with different computational capabilities, memories with different capacity/bandwidth characteristics (e.g., HBM vs. DDR), or networks with different bandwidth/latency characteristics (e.g., InfiniBand vs. NVLink).
Next, we describe how individual performance models are constructed and tied together in \TheName.
}
\vspace{-3mm}
\subsubsection{Compute delay estimation}
\label{sec:method:compute-delay}

To produce compute-delay estimations independent of any compute node's microarchitectural details, we
employ a roofline model~\cite{Georg:roofline_application, samuel:roofline}.
In each case, the compute node of interest is represented by its peak performance (${perf}_{peak}$ in $GFLOPS$) and memory bandwidth.

\begin{figure}[t]
    \centering    
    \includegraphics[width=.8\columnwidth]{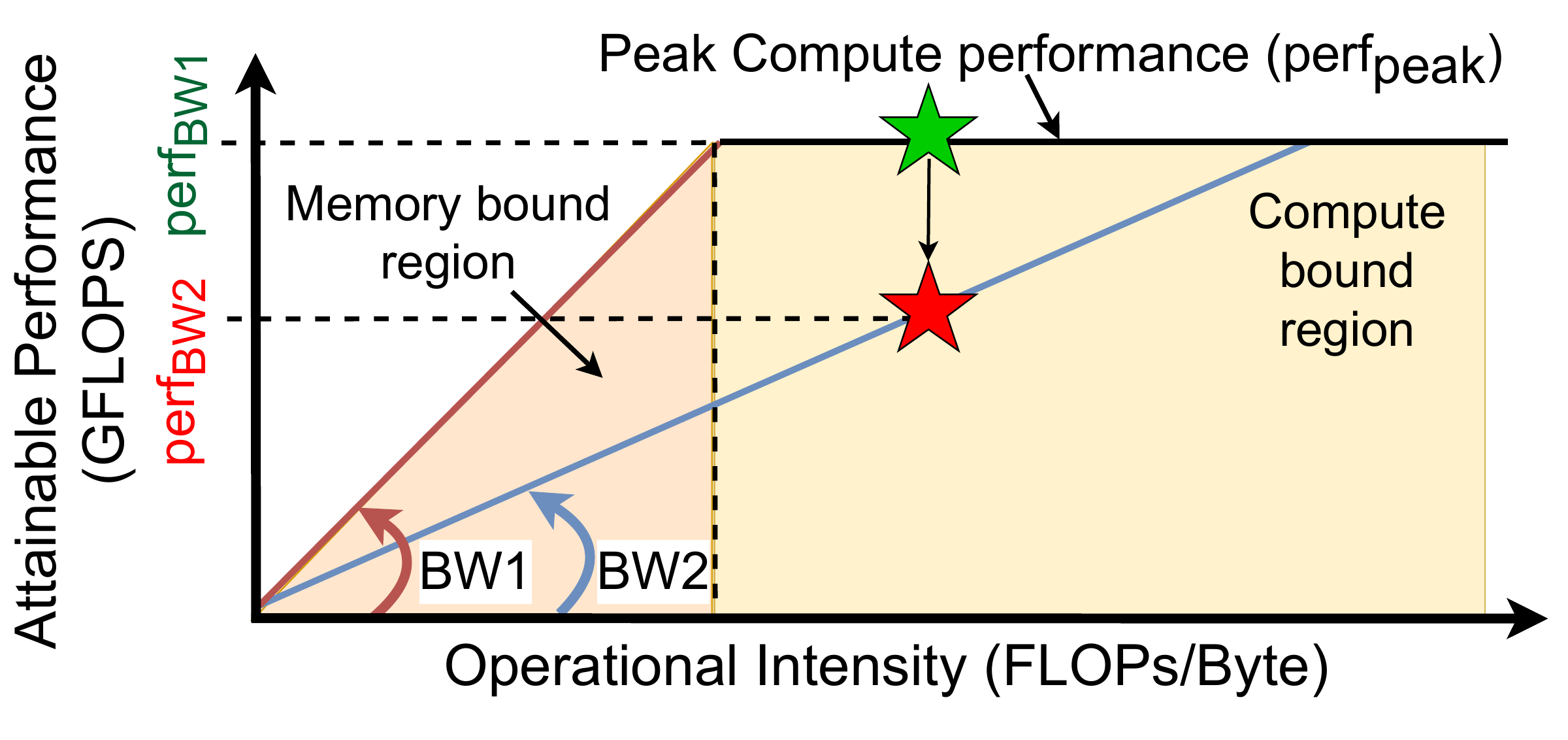}
    \vspace{-3mm}
    \caption{Roofline model. Attainable performance shifts for the same OI, depending on available memory bandwidth.
    }
    \label{fig:roofline}
    \vspace{-4mm}
\end{figure}

\cref{fig:roofline} shows a roofline model, which consists of a compute-bound region ({under the flat line}) and a memory-bound region ({under the slanted line}).
The flat line is dictated by the target node's ${perf}_{peak}$.
The slanted line's slope denotes the compute node's available memory bandwidth $BW_{mem}$($GB/s$).
In \TheName's design space exploration, a change in the available memory bandwidth changes the roofline's slope and, correspondingly, the intersection point with the ${perf}_{peak}$ horizontal line shifts. 

For each training phase, a workload layer's operational intensity ($OI$) is calculated as
\vspace{-1mm}
\begin{equation}
    OI\,(FLOPs/byte) = \frac{\textit{\# of floating-point operations}} {\textit{memory traffic}}
    \label{eqn:oi}
\end{equation}

\noindent where \textit{$\textit{memory traffic}$} is the total number of bytes moved between the memory and processor for operand and result matrices of the layer.
\cref{sec:method:data-movement-estimation} elaborates on our estimation of memory traffic generated by each layer.
Based on the target compute node's characteristics, a workload layer's OI may place it under the roofline's compute-bound or memory-bound region.
The maximum compute performance %
achieved for a layer $i$ with $OI_i$ is 
$\textit{perf}_{max}\,(GFLOPS) = min\left\{{perf}_{peak},\,OI_i\times BW_{mem}\right\}$.

Using ${perf}_{peak}$ and $BW_{mem}$ as knobs in our methodology, we estimate their impact on the maximum attainable performance $\textit{perf}_{max}$ for each workload layer, and thereby the compute delay for each layer $i$ in each training phase as
\begin{equation}
    compute\ delay_i\,(s) = \frac{\textit{\# of floating-point operations}}{\textit{perf}_{max}}
\end{equation}
The total \textit{compute} delay %
for one training iteration is the sum of compute delays for the forward pass, backward pass, and weight update for each of the model's layers.

\revision{The roofline model makes our workflow fast and versatile, and while not enough to estimate absolute performance, it captures performance trends. 
Despite having limitations, as a first-order model, the roofline model has still been used widely in a broad range of prior work \cite{roofline:ding:PMBS, roofline:ibrahim:springer, roofline:li:IPDPS, roofline:miao:ISCC,  Georg:roofline_application, roofline:wang:ICBAIE, roofline:wang:DLS,  samuel:roofline, roofline:yang:arxiv, ridgeline, roofline:yang:springer} due to its versatility and utility in quickly and correctly highlighting general \textit{performance trends}, given the general compute and memory capabilities of a compute node.
\TheName as a general methodology is not limited to roofline and its modular design allows plugging in more detailed compute models, such as runtimes captured from real GPUs or simulated accelerators modeled in appropriate tools, 
{like GPGPU-Sim \cite{accelsim}, or ScaleSim \cite{samajdar:scalesim}.
However, in this paper we use the more general roofline model to focus on the methodology's utility and decouple the results of the conducted case studies from any specific compute unit's microarchitectural characteristics.}}

\subsubsection{Memory traffic estimation}
\label{sec:method:data-movement-estimation}

Memory traffic is the cumulative number of bytes transferred between the main memory and compute unit while performing the desired functionality.
For a hypothetical compute node with infinite on-chip buffer space, all operands can be fetched exactly once from the memory, resulting in a very high OI (cf. \cref{eqn:oi}).
However, every realistic compute unit's limited on-chip buffer space can only hold a limited set of data operands.
Therefore, a layer's matrix operands must usually be fetched multiple times from the memory to complete the required operations, thereby lowering the resulting OI. 
We construct a linear model to better estimate the memory traffic for a GEMM operation on a compute node with an on-chip buffer \revision{of configurable} size.

Consider a GEMM operation between two matrices of $U$ and $V$ bytes, generating an output matrix of $W$ bytes.
We assume one of the input operands is tiled to fit in the on-chip buffer, and the other operand/output are streamed in/out of the compute node, respectively.
For an on-chip buffer size of $S$ bytes, we estimate the memory traffic (in bytes) as 
$min\left\{\Psi_{1},\Psi_{2}\right\} + W$, 
where $\Psi_{1} = \lceil U/S\rceil \times V + U$ and $\Psi_{2} = \lceil V/S \rceil \times U + V$.
\revision{In practice, for $U$ and $V >> S$, tiling the smaller operand results in less data movement (e.g., if $U<V$, $\Psi_{1}$ is the tiling method of choice, resulting in about $V-U$ less data movement).}

An additional important architectural design knob we want to investigate with \TheName is that of memory expansion, whereby the compute unit's local memory (LM) is enhanced with a secondary level of memory, which we refer to as expanded memory (EM). 
Such a setting can be enabled by allowing the compute unit to directly access its host CPU's memory, or by physically attaching additional memory over CXL~\cite{cxl}, photonic links, or other (current or future) technology.
Investigating such an option using CXL-attached memory (which offers considerably higher bandwidth per pin than DDR) for DL training is of particularly high relevance given the growing interest in deploying it as a new memory hierarchy component~\cite{cxl-pooling,pond, tpp}.

To investigate this system design option, we consider per-node memory expansion, with the available bandwidth as our sensitivity analysis knob. 
To model performance with such a hybrid memory system, we instrument our roofline model with the new memory system's effective memory bandwidth (${bw}_{hybrid}$), which depends on the fraction of data accessed from local/expanded memory ($data_{LM}/data_{EM}$) at the local/expanded memory's bandwidth ($bw_{LM}/bw_{EM}$).
We estimate the hybrid memory system's effective bandwidth as: 
\begin{equation} \label{eq:ext-mem-bandwidth}
    bw_{hybrid}= \frac{total\_data\_accessed}{\frac{data_{local}}{bw_{LM}}\,+\,\frac{data_{EM}}{bw_{EM}}}
    \vspace{-0.5em}
\end{equation}

To illustrate, accessing 240GB of data in a hybrid memory system with 80GB of LM, $bw_{LM}=2TB/s$, and $bw_{EM}=1TB/s$ results in $bw_{hybrid}=1.2TB/s$. %
Using \cref{eq:ext-mem-bandwidth}, we can determine the cluster's performance as a function of the bandwidth offered by the hypothetical memory expansion technique used.

\subsubsection{Communication delay estimation}
\label{sec:method:communication-delay}

During the training process, nodes continuously exchange data.
Their communication delay is dictated by the total communication volume, the aggregate network bandwidth available between the nodes, and the dynamic network utilization.
In addition, depending on the training phase, the communication among the nodes may be  blocking or non-blocking.
In the forward-pass and input-gradient phases, communication is blocking along the model-parallel dimension; in the weight-gradient phase, communication is non-blocking across the data-parallel dimension.
Blocking communication falls on the critical path of a training phase, %
while non-blocking communication can be (partially) overlapped with compute, thus ameliorating its impact on resulting training time.
The combination of data movement volume, available network bandwidth, communication type, and concurrently performed computation, dictates how much of the occurring communication is exposed, affecting training time.
Ultimately, for each layer, the total exposed communication delay determines whether the layer is compute- or communication-bound on a given system configuration.

\subsubsection{Total training time estimation}
\label{sec:method:total-training-time}
Finally, we combine the per-layer compute delays (\cref{sec:method:compute-delay}) with each layer's corresponding communication characteristics (data volume and communication type---\cref{sec:method:communication-delay}) to determine the degree of computation/communication overlap and derive the training time per iteration and, by extension, total training time. 

Overall, \TheName comprises an iterative training time estimation process as shown in \cref{fig:method-overview}.
For each model, different training strategies are considered (cf. \cref{sec:method:training-strategy}) and the training time is estimated as described in \cref{sec:method:training-time-estimation}.
The process is repeated for different cluster sizes, and compute, network, and memory parameters to guide the user's selection of parallelization strategy and cluster resource provisioning to optimize for the target metric of merit---raw training performance, or training efficiency (i.e., training time relative to resources deployed).

\section{\TheName Implementation}
\label{sec:instantiation}

\begin{figure}[t]
    \centering
    \includegraphics[width=\columnwidth]{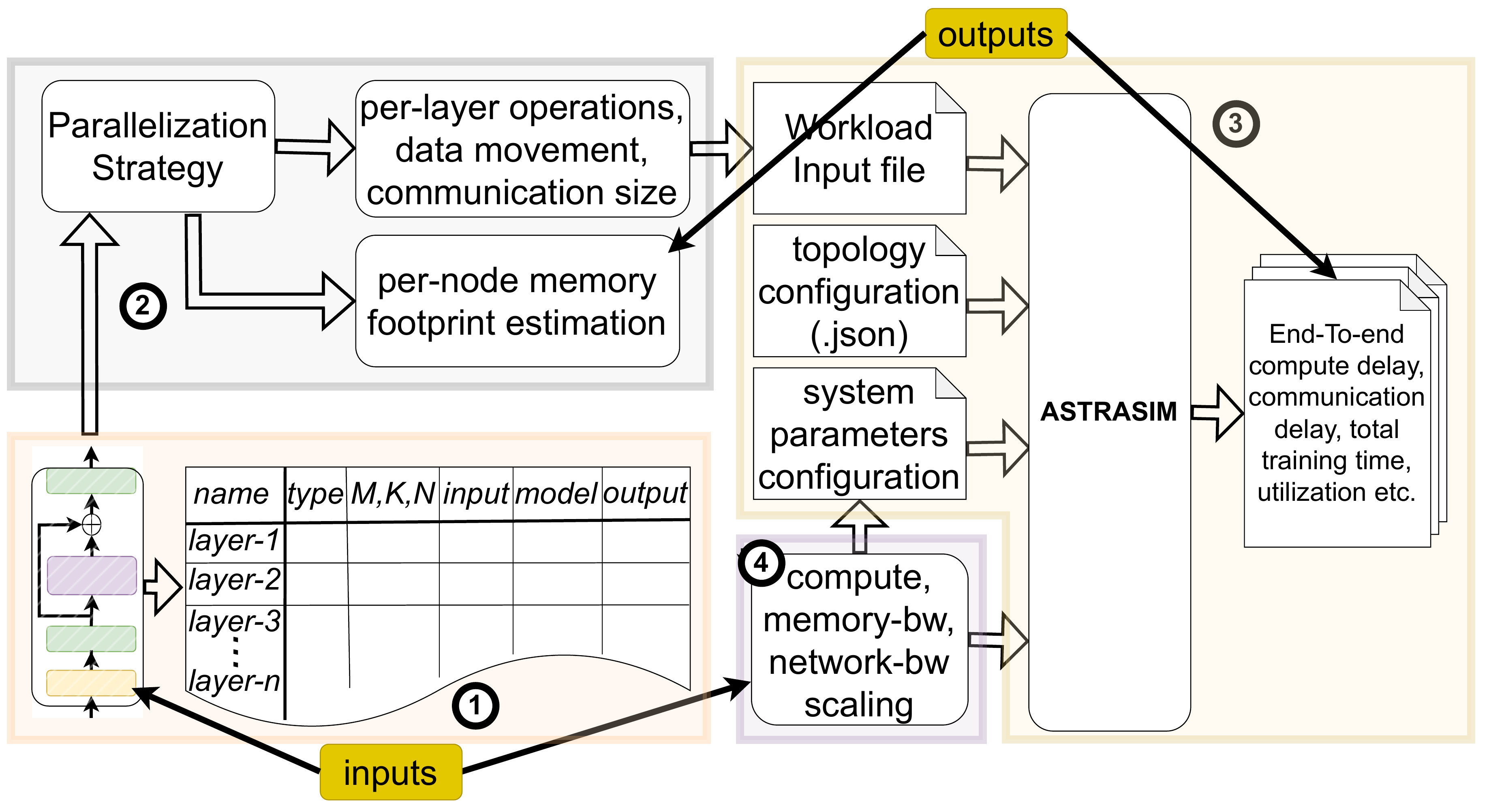}
     \vspace{-6mm}
    \caption{\TheName implementation and workflow.}
    \label{fig:COMET-instance}
    \vspace{-5mm}
\end{figure}

\cref{fig:COMET-instance} shows the implementation of our toolchain implementing the \TheName methodology. %
\revision{The toolchain's frontend, in steps \TextCircled{1} and \TextCircled{2}, generates parameters for the per-layer computation and communication requirements.}
These parameters are then fed into the toolchain's backend, which models the resulting performance of the training task, as a function of the target cluster's  configuration (steps \TextCircled{3} and \TextCircled{4}).

\subsection{DL Model Analysis}
\label{impl:step1}

In step \TextCircled{1}, we analyze the DL model of interest and break it down into its  layers.
Each layer is represented as a sequence of GEMMs of input activations, model parameters and resulting output activation matrices.
The size of each matrix dimension M, K, N is derived from the model's hyper-parameters and batch size.
Depending on the model type, a set of fixed independent hyper-parameters form the model's signature. 
For example, in case of Transformers, hidden-dimension, \# layer-stacks, and \# attention-heads characterize the model size.
Then, based on the derived GEMM dimensions, we compute the total number of model and activation parameters required for each layer.
In addition, the size of each operand matrix (i.e., input activations, model parameters, and output matrices) per node is affected by the MP/DP degree.

\subsection{Parallelization Strategy}
\label{impl:step2}

In step \TextCircled{2}, we select a parallelization strategy for the given workload, generate the corresponding workload input file and feed it to the performance simulator to estimate the distributed training time.
In addition, the workload's required memory footprint is computed by aggregating model parameters, optimizer states, gradients, residual states, and checkpoint-activations.
The workload input file must describe the characteristics of each layer of the workload, which includes  the number of floating-point operations, data volume (in bytes) moved between memory and compute, communication collective, and communication volume (in bytes).
As per ZeRO-Infinity, we compute the total number of operations and matrix operands required in both forward (FP) and backward (i.e., input gradient (IG) and weight gradient (WG)) training phases.

As described in \cref{sec:method:data-movement-estimation}, we estimate the bytes transferred between the processor and main memory for each layer in each training phase.
Similarly, we estimate the total number of operations and matrix operands required for each layer in each training phase.
Based on the parallelization scheme used for the workload, we also determine the communication collective and compute the communication volume required per layer in each training phase, as well as the required per-node memory footprint to fit the model states and working dataset.

We use ZeRO-DP (os+g) \cite{rajbhandari:zeroDP}, a.k.a. ZeRO-2, to derive the per-node memory footprint of model-states. %
ZeRO-2 avoids replication of optimizer states and gradients on each node and distributes them across the data-parallel dimension to reduce the per-node memory footprint, while avoiding additional communication overhead.
For residual states, we estimate the memory footprint as $\textit{\# activation-parameters} \times \textit{2 bytes}$ assuming fp16 activation parameters.
We exclude the memory required for checkpoint activations in our per-node memory footprint estimate.
Typically, in large models such as \TheModel, the memory footprint required to store the checkpoint activations is significantly larger than the intermediate activations and hence are offloaded to host memory.
Therefore, during the training process, we only consider the Activation Working Memory~\cite{rajbhandari:zero-infinity}, which is the memory required to hold the intermediate activations between two consecutive checkpoints.

\begin{figure}[t]
    \centering
    \includegraphics[width=\columnwidth]{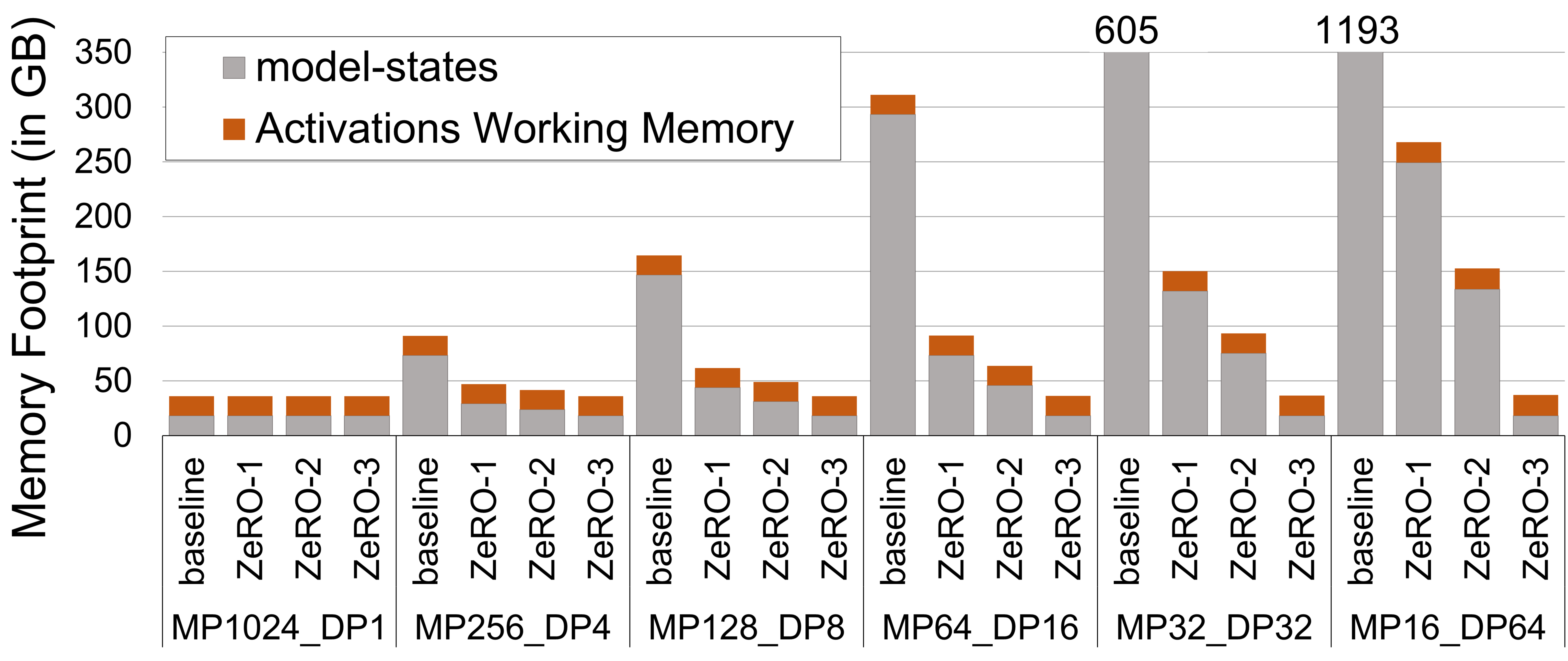}
    \caption{Per-node memory footprint for Transformer-1T model with baseline and different ZeRO-DP stages. }
    \label{fig:ZeRO-DP-comparison}
    \vspace{-3mm}
\end{figure}

As explained in \cref{sec:method:training-strategy}, the parallelization strategy affects the required memory footprint per node. To illustrate with a concrete use case, \cref{fig:ZeRO-DP-comparison} shows how the per-node memory footprint requirement changes for a \TheModel model on a fixed cluster size of 1024 nodes, as a function of different stages of ZeRO and a decreasing MP degree (the invariant being $DP\times MP=1024$). 
In baseline (i.e., no ZeRO optimizations), the model footprint per node increases exponentially as the MP degree reduces.
The same trend holds even with memory optimizations like ZeRO-2: although the growth is slower, the model footprint per node eventually exceeds the typical memory capacity of a single device, highlighting the value of MP to enable in-memory training of huge models.
Among all the ZeRO-DP optimizations, ZeRO-3 stands out as it provides the lowest memory footprint per node and remains unaffected by MP reduction. 
However, ZeRO-3 incurs a $1.5\times$ communication overhead compared to the baseline.
Other approaches such as ZeRO-Offload and ZeRO-Infinity offload data and compute to the host machine resources to reduce the memory capacity pressure on accelerator nodes.

\subsection{Total Training Time Estimation}
\label{impl:step3}

{As shown in \autoref{fig:COMET-instance} step \TextCircled{3}, \TheName plugs into a cost model for training time estimation.}
We use the \astrasim simulator \cite{rashidi:astrasim, won:astrasim2} {for this purpose}.
\astrasim is a discrete event-based simulator {developed by Meta, Intel and Georgia Tech} that can simulate distributed training for a variety of DL workloads.
It accepts a workload configuration, topology description and system parameter file to simulate the distributed DL training on a target cluster, and outputs the end-to-end training time breakdown and resource utilization. 

At a high level, \astrasim consists of a workload, system, and network layer.
The workload layer is responsible for instantiating a model and scheduling training loops for simulation.
The system layer provides the mechanisms for collective primitives and scheduling of communication tasks, {similar to collective communication libraries like NCCL~\cite{nccl}}.
The network layer provides the topology interface via network APIs to support a fast analytical model~\cite{won:astrasim2} or detailed network simulation using Garnet \cite{agarwal:garnet} and NS3 \cite{riley:NS3}.

An \astrasim workload configuration file describes the DL workload to be simulated, and consists of the compute time, communication collective, and the collective size for each layer of the workload.
For each layer, \astrasim schedules the communication collectives and \textit{overlaps} the communication delay with compute delay to estimate the total training time.
\astrasim's analytical network backend---used in our current \TheName implementation---estimates the communication delay of an event based on the topology and network bandwidths specified in the configuration files.
\astrasim supports multiple  collective communication primitives and scheduling algorithms. The symmetric network topology of distributed training platforms and topology-aware collective communication algorithms minimize the network congestion, enabling the analytical network backend to accurately model the communication overhead \cite{ khan:2022, rashidi:2021, rashidi:themis,won:astrasim2}. 
{\astrasim's runtime projections for 8--16 node clusters has been validated against real systems to be within 5\% difference~\cite{won:astrasim2}}.

{For \TheName, we chose \astrasim as the cost model for training time estimation given its modular architecture for plug-and-play compute and communication models, and implementations of diverse collective communication algorithms and scheduling strategies. We integrated our roofline and data movement models (\cref{sec:method:compute-delay,sec:method:data-movement-estimation}) in  \astrasim to enable modeling a range of compute units with hybrid memories (LM and EM). 
Using our added models and the data provided in the workload input file (operations and data size per layer), the compute delay per layer is estimated. }
\astrasim  uses the compute delay, communication collectives and volume, the network topology, and system parameters provided to perform a training simulation and generates an end-to-end per-layer training-time breakdown for each training phase.
It reports the compute and exposed (i.e., non-overlapped) communication times for each layer in the FP, IG, and WG phases. 

For our design space exploration of parallelization strategies on a cluster of size $N$, we sweep the degree of MP and DP such that always $MP \times DP = N
$. 
This emulates the effective increase in per-node memory capacity as MP decreases in favor of DP, as demonstrated in \cref{fig:ZeRO-DP-comparison}, and generates the corresponding workload input file for each (MP, DP) combination.

\subsection{Cluster Parameter Reconfiguration}
\label{impl:step4}

Finally, step \TextCircled{4} provides a set of knobs to perform sensitivity analysis by varying the key component parameters, which include the network topology, compute capability, as well as memory and network bandwidth and latency. 

\subsection{Iterative Modeling for Design Space Exploration}

By iterating through steps \TextCircled{2} to \TextCircled{4} (\cref{impl:step2}--\cref{impl:step4}), we obtain the resulting training time for different system configurations and hardware parameters to identify the best combination of parallelization strategy and cluster resources.
The target optimization metric may be raw performance or cost efficiency (i.e., performance relative to cluster's provisioned resources).
\section{Case Studies}
\label{sec:case-studies}
We now leverage \TheName to evaluate cluster design decisions across multiple dimensions (cluster size and network, per-node memory and compute capability) in the context of large-scale Transformer and DLRM training.
We first describe the models and the cluster we model as our baseline system (\cref{sec:baseline}). 
We then evaluate the impact of different cluster configuration parameters using Transformer (\cref{sec:transformer-eval}) and DLRM (\cref{sec:dlrm-eval}) models.
\cref{sec:cluster-comparison} employs \TheName to compare a range of different clusters on a range of models.
Finally, \cref{sec:versatility} concludes with a summarizing overview of \TheName's versatility and  quantification of the tool's key strength of \textit{speed}, enabling rapid design space explorations for distributed training tasks on large clusters.

\begin{figure}
    \centering
    \includegraphics[width=\columnwidth]{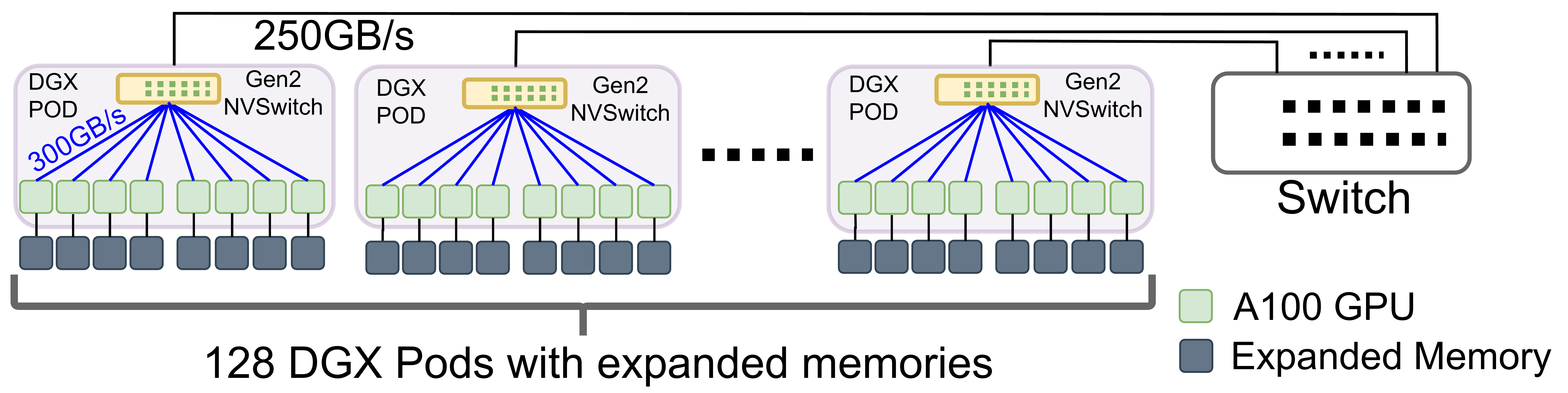}
    \caption{Cluster of 1024 A100 GPUs with expanded memories, grouped in 128 8-GPU pods. Link bandwidth is per direction. %
    }
    \label{fig:DGX}
\end{figure}
\begin{table}
    \centering
    \footnotesize
    \caption{Baseline NVIDIA DGX A100 system parameters.}
    \label{tab:baseline-param}
    \scalebox{0.94}{
        \begin{tabular}{ll}
        \toprule
        \multicolumn{2}{l}{\textbf{Single-node Parameters (NVIDIA A100 GPU)}}  \\
        \midrule
        Peak Performance ($\textit{perf}_{peak}$) & 624 TFLOPS (fp16)\\
        {Local Memory Capacity / Bandwidth}& 80 GB / 2039 GB/s\\
        On-chip SRAM size & 40 MB \\
        \toprule
        \textbf{Cluster Parameters}\\
        \midrule
        Compute Pod & NVIDIA A100 DGX (8-GPU) \\
        Cluster Size & 1024 nodes (128 pods $\times$ 8 GPUs) \\
        Intra-pod Network BW per GPU & 300 GB/s / direction (NVLink Gen-3)\\
        Inter-pod Network BW per GPU & 31.25 GB/s / direction (InfiniBand) \\
        Physical Topology & Hierarchical Switch\\
        Collectives Implementation & Logical Ring\\
    \bottomrule
    \end{tabular}
    }
    \vspace{-3mm}
\end{table}

\subsection{Baseline Evaluation Setup and Workloads}
\label{sec:baseline}

\cref{fig:DGX} visualizes our baseline 1k-node DGX A100 \cite{nvidia:dgxa100} cluster with expanded memories and \cref{tab:baseline-param} summarizes the parameters used to model it in \astrasim. 
We evaluate training performance for Transformer and DLRM models, which represent the largest models currently deployed.
Our evaluation predominantly focuses on the Transformer model due to its higher complexity and broader range of (MP, DP) training strategies (\cref{sec:transformer-eval}).
Due to space constraints, we present a small subset of DLRM evaluation results in \cref{sec:dlrm-eval}.

\subsubsection {Transformer Model}
\label{sec:transformer-details}
Transformer-based language models are huge natural language processing models used extensively in language modeling, machine translation, text summarization, AI chatbots (e.g., ChatGPT), etc.
Latest Transformer models comprise up to trillion parameters
and must be trained over hundreds of high-end processing nodes in a distributed manner using multiple levels of parallelization \cite{narayanan:megatronlm,DBLP:shoeybi:megatronlm}, despite advanced  techniques  employed reduce their required memory footprint \cite{Prakhar:embedding-compression,rajbhandari:zeroDP, rajbhandari:zero-infinity}. %

A Transformer model comprises a stack of multiple encoder and decoder structures, each composed of multi-head attention layers followed by fully connected feed-forward and residual layers \cite{vaswani:attention}. 
Each of the encoder and decoder stack takes input and output embeddings as inputs which map a sequence of symbols into a continuous representation.
We model the Transformer-1T architecture and hybrid model \& data parallelism approach as described in Megatron-LM \cite{DBLP:shoeybi:megatronlm}. %
\cref{tab:transformer-layers} breaks down the \TheModel model into its layers, and summarizes each layer's type and dimensions.
\begin{table}[t]
    \centering
    \small
    \caption{Transformer model layers and dimensions. }
    \label{tab:transformer-layers}
    \scalebox{0.65}{
    \begin{tabular}{lccccc}
    \toprule
        \multirow{2}{*}{\bf{Layer} } & 
        \multirow{2}{*}{ \bf{Type}  } & 
        \multirow{2}{*}{\bf{\#Stacks}} &  
        \multicolumn{3}{c}{\bf{GEMM Dimensions}}\\ \cline{4-6} & & & M & K & N \\
    \toprule
        Input Embedding & Table Look-up & 1 & $b\times seq$ & $sub\_vocab$ & $d_{model}$ \\
        Layer Norm & Element-Wise Mult & N & $b\times seq$ & 1 & $d_{model}$ \\
        Query Projection ($Q_{i}$) & GEMM & N & $b\times seq$ & $d_{model}$ & $h\times d_{k}$\\
        Key Projection ($K_{i}$) & GEMM & N & $b\times seq$ & $d_{model}$ & $h\times d_{k}$\\
        Value Projection ($V_{i}$) & GEMM & N & $b\times seq$ & $d_{model}$ & $h\times d_{v}$\\
        $U_{i} = softmax(Q_{i}K^{T}_{i} / \sqrt{d_{k}})$ & GEMM & N & $b\times seq$ & $h\times d_{k}$ & $b\times seq$\\
        $Y_{i} = U_{i}V_{i}$ & GEMM & N & $b\times seq$ & $b\times seq$ & $h\times d_{v}$\\
        concat($Z_{i}=Y_{i}B_{i},..,Z_{n}$) & GEMM & N & $b\times seq$ & $h\times d_{v}$ & $d_{model}$\\
        Residual Addition & Element-Wise Add & N & $b\times seq$ & 1 & $d_{model}$\\
        Layer Norm & Element-Wise Mult & N & $b\times seq$ & 1 & $d_{model}$\\
        $Y_{i}=GeLU(XA_{i}+k)$ & GEMM & N & $b\times seq$ & $d_{model}$ & $sub\_{\textit{ff}}$\\
        $Z_{i}=Y_{i}B_{i}+k$ & GEMM & N & $b\times seq$ & $sub\_{\textit{ff}}$ & $h\times d_{model}$ \\
        Residual Addition & Element-Wise Add & N & $b\times seq$ & 1 & $h\times d_{v}$ \\
        Output Embedding & Table update & 1 & $b\times seq$ & $d_{model}$ & $sub\_vocab$ \\
    \bottomrule
    \multicolumn{6}{l}{   \textbf{Legend} - $d_{model}$: hidden dimension, $h$: \# of attention heads,  $b$: mini-batch size, $seq$: sequence length,}\\ \multicolumn{6}{l}{$d_{k}$/$d_{v}$: key/value tensor dimension per attention head, $sub_{\textit{ff}}$: portion of MLP layer per MP node,}\\
    \multicolumn{6}{l}{$sub\_vocab$: chunk of total vocabulary per MP node}\\
    \end{tabular}    
    }
    \vspace{-5mm}
\end{table}

\begin{figure*}[t]
    \centering 
    \captionsetup[subfloat]{captionskip=-2pt}   	

    \subfloat[Breakdown of comp./comm. time \& per-node memory footprint.]{
    \includegraphics[width=\columnwidth]{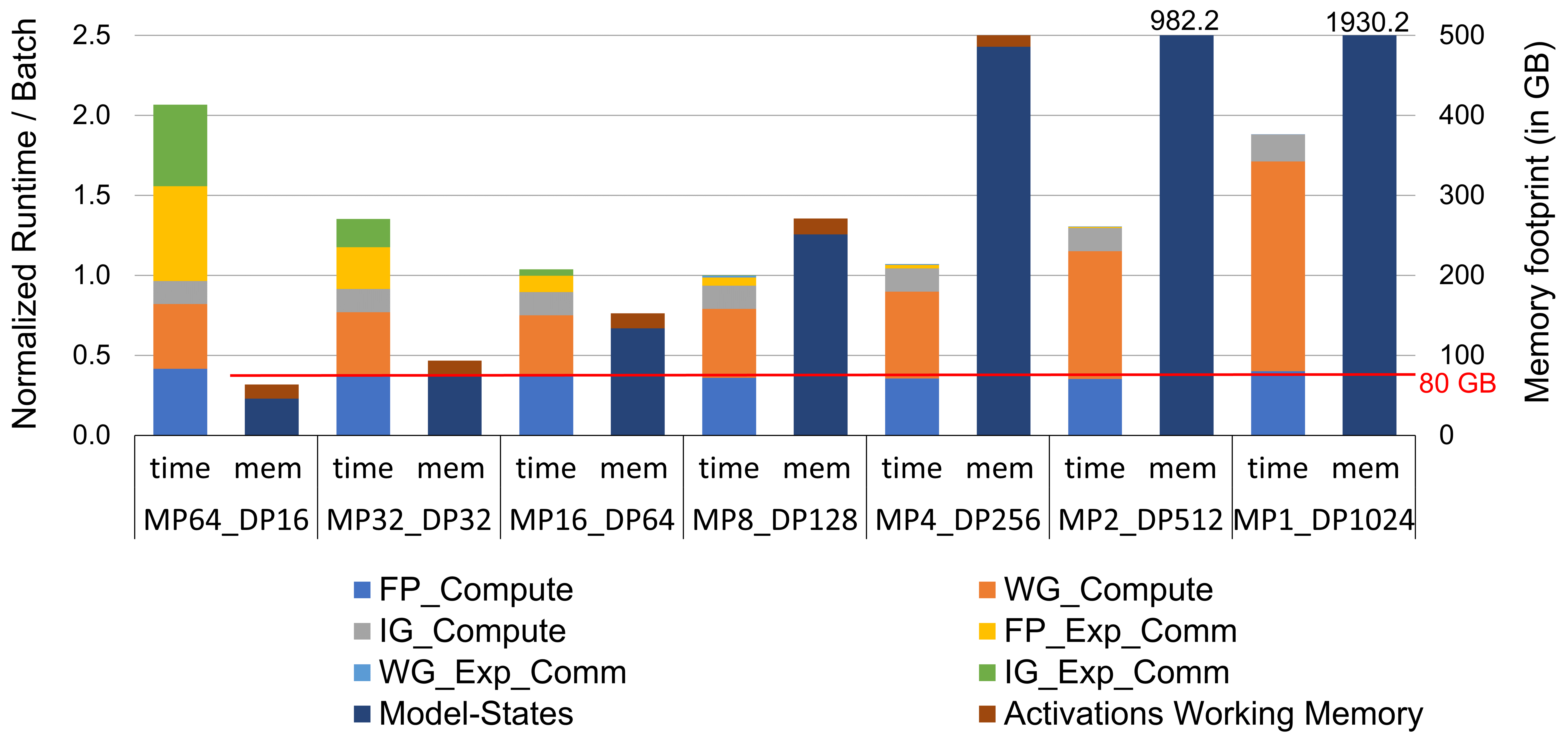}
    \label{fig:baseline-TT-memFP-breakdown}
    }
    \subfloat[Exposed communication to compute ratio.]{
    \includegraphics[width=\columnwidth]{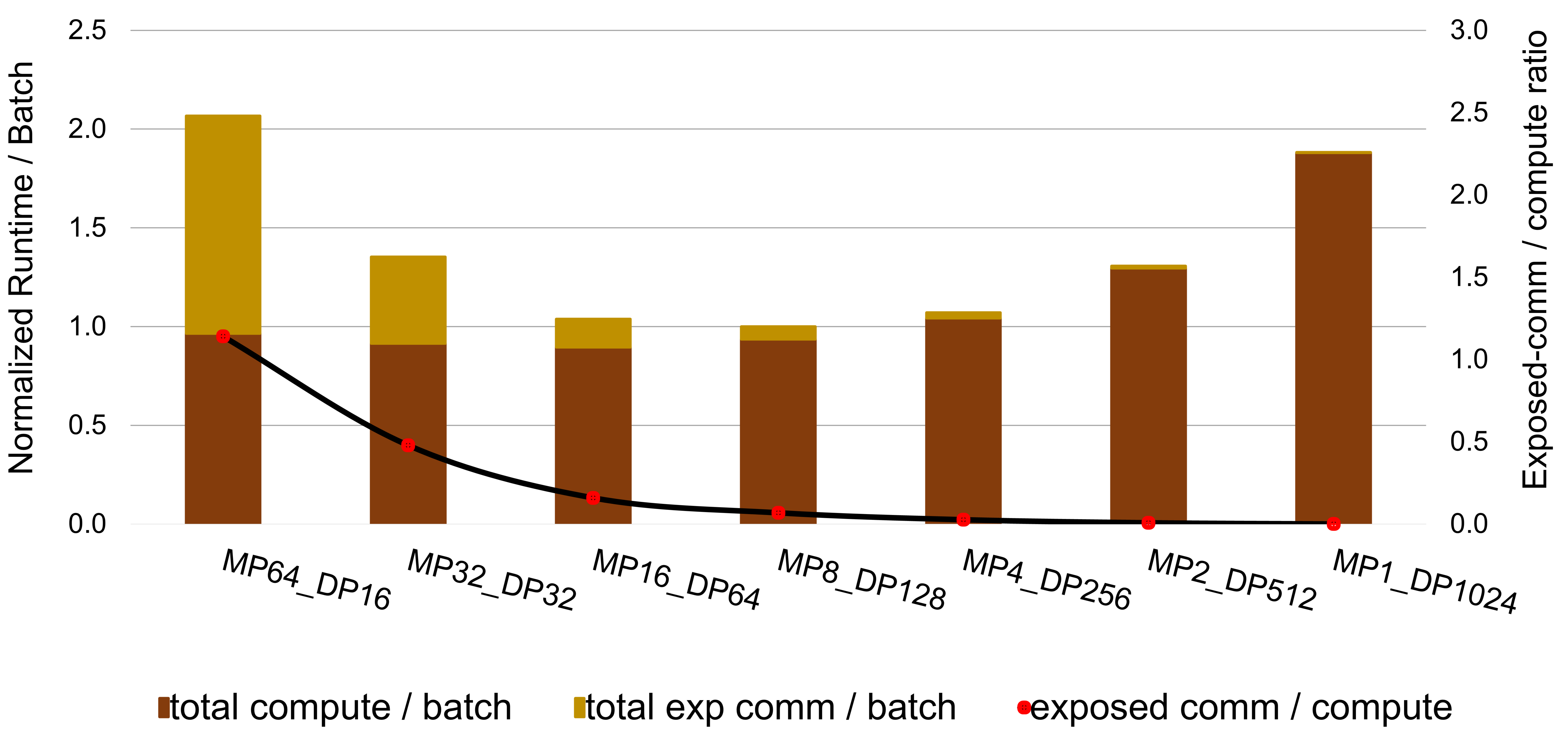}    
    \label{fig:exp-comm-to-compute}
    }
    \caption{Training runtime of \TheModel with varying MP/DP degree (norm. to best-performing MP8\_DP128 config.).     }
\end{figure*}

\subsubsection {DLRM}
\label{sec:dlrm-details}

DLRMs are among the largest deep learning models used widely for generating personalized content \cite{covington:2016, gupta:2020, naumov:2020} (e.g., movie recommendations \cite{gomez2016, Koren:2009}, e-commerce catalogs \cite{smith:2017, wang:2019, zhou:2019}), comprising up to trillions of parameters.
DLRM model contains large embedding tables that capture the latent space of user and product feature interaction. 
In each DLRM, inputs are classified into sparse and dense features which represent the categorical data and continuous features.
While sparse features are used to look up the embedding tables, dense features are processed using a bottom stack of MLP layers.
The result of embedding lookup and output of bottom MLP layers are combined and given as input to the top MLP layers which finally predict the probability of an event (such as click-rate) occurrence.
We model the DLRM architecture and parallelization strategy as described by Rashidi et al. \cite{rashidi:2020}.

Unlike Transformer models that can be trained using a wide range of model versus data parallelism configuration points, the training structure for DLRMs is more rigid.
DLRM follows a hybrid parallelization strategy where the bottom embedding layers are sharded across multiple nodes and performs an all-to-all communication during forward and backward propagation while the MLP layers are replicated on each node in a data parallel fashion and perform an all-reduce during backward propagation.
DLRM therefore does not offer the same MP/DP configuration knob we sweep for Transformer models.

\subsection{Transformer Evaluation}
\label{sec:transformer-eval}

A Transformer model can be trained via a wide range of parallelization strategies. We first evaluate the impact of different (MP, DP) parallelization strategies on that cluster's performance and per-node memory requirements (\cref{sec:evaluation:memory-capacity}).
We then demonstrate the impact of individually scaling the cluster's per-node memory system (\cref{sec:transformer-memory}), per-node compute capability (\cref{sec:transformer-compute}),  and network (\cref{sec:transformer-networking}).

\subsubsection{Parallelization Strategy Impact on Performance and Memory Requirements}
\label{sec:evaluation:memory-capacity}
We begin by %
sweeping (MP, DP) under the invariant $MP\times DP=1024$.
In this section, training time estimations ignore per-GPU memory capacity constraints, assuming infinite per-node memory capacity accessible at the baseline system's peak memory bandwidth.

\cref{fig:baseline-TT-memFP-breakdown} shows the training time breakdown and corresponding per-node memory footprint for several (MP, DP) configurations, assuming a constant memory bandwidth of 2039GB/s, irrespective of capacity.
The total training time is a combination of compute delays and exposed communication delays (cf. \cref{sec:method:total-training-time}).
The compute and exposed communication time is broken down into three components to indicate the three main phases in a training iteration: forward pass (FP), input gradients (IG), and weight gradients (WG).
As MP decreases in favor of more DP groups, the required memory footprint per node increases; therefore, fitting the model in our baseline GPU's 80GB memory requires an MP degree of 64 or higher.

Memory capacity requirements aside, the best-performing configuration is MP8\_DP128, where the exposed communication in FP and IG phases, and the compute delay exhibit their lowest values.
Note that WG communication (\textit{WG\_Exp\_Comm}) is fully overlapped by the WG compute (\textit{WG\_Compute}) in every configuration, hence not visible.
The configurations left of MP8\_DP128 (higher MP) are communication bound due to the exposed blocking communication patterns in FP and IG phases across the MP dimension.
In contrast, for configurations right of MP8\_DP128 (lower MP), the effective model footprint per node grows as the model is distributed across fewer nodes per DPU, and hence becomes more memory bound resulting in higher compute delays dominating runtime, while barely any communication is exposed.

\cref{fig:exp-comm-to-compute} better highlights the changing balance between compute and exposed communication time for the same (MP, DP) range.
Under high MP degrees (e.g., MP64\_DP16), training time is dominated by exposed communication time. 
As MP decreases in favor of increasing DP, the fraction of runtime spent on communication becomes negligible from MP8 onwards.
\textit{MP8\_DP128 is the optimal configuration because it strikes the best balance, effectively overlapping communication delays with compute, without getting into a memory-bandwidth-bound region that causes drastic compute time increase.}

\medskip
\subsubsection{Effect of Memory System Design}
\label{sec:transformer-memory}

Based on \cref{sec:evaluation:memory-capacity}'s results,  the best-performing MP8\_DP128 configuration requires $\sim$250GB of memory to fit the model, exceeding the 80GB the NVIDIA A100 GPU baseline's per-node memory capacity by $3\times$.
The best-performing configuration achievable under the 80GB memory constraint is MP64\_DP16.
The required capacity for MP8\_DP128 could be achieved with a hybrid memory system, complementing the GPU's HBM with a secondary DRAM-based memory that offers additional capacity, albeit accessible at lower bandwidth.
As mentioned in \cref{sec:method:data-movement-estimation}, such additional capacity could be provided by allowing the GPU to access its host CPU's memory, or by attaching additional memory to the GPU over CXL \cite{cxl} or other technology.

\begin{figure}
    \centering    
    \includegraphics[width=\columnwidth]%
    {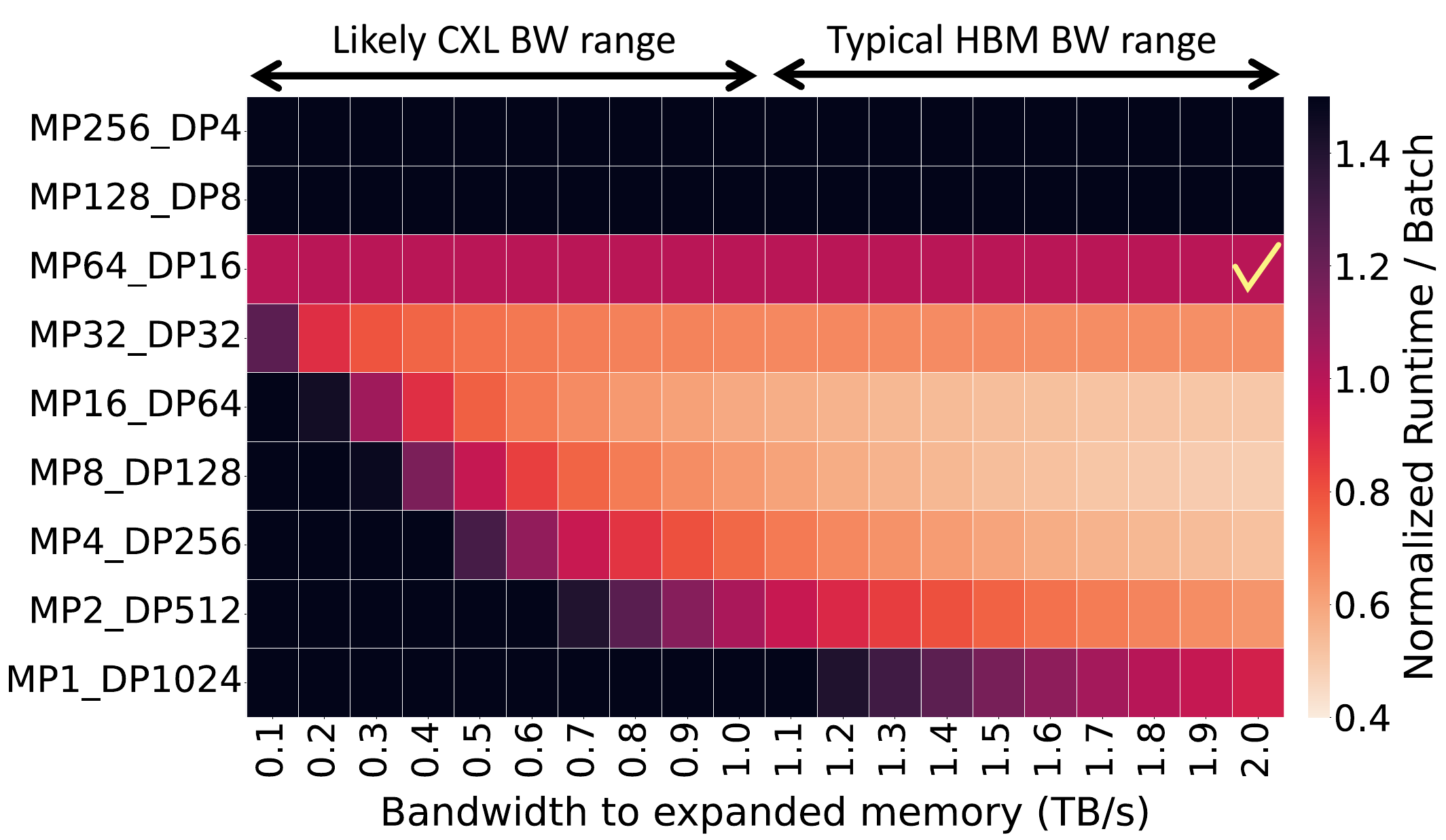}
    \caption{Effect of memory bandwidth availability to the expanded memory system. The check-mark indicates the baseline configuration to which all the other values are normalized. %
    }
    \label{fig:transformer-mem-bw-scaling}
    \vspace{-5mm}
\end{figure}

\cref{fig:transformer-mem-bw-scaling} shows the performance results normalized to MP64\_DP16, the best-performing configuration of in-memory distributed training that is feasible without memory expansion %
(cf. \cref{fig:baseline-TT-memFP-breakdown}).
The heatmap's x-axis shows the bandwidth to expanded memory, while the varying (MP, DP) degree on the y-axis is a proxy for the required capacity of that expanded memory (see memory requirements in \cref{fig:baseline-TT-memFP-breakdown}). 
MP64\_DP16 and configurations with higher MP remain unaffected by the expanded memory's bandwidth, as the dataset entirely fits in each node's local memory, and configurations with MP higher than 256 are omitted, as they perform strictly worse. 

\textit{The heatmap guides system architects in determining what memory expansion technology can be used to boost a cluster's training performance, revealing the range of expanded memory characteristics that allow building a cluster with lower training time than the MP64\_DP16 baseline.
Conversely, the data can also be leveraged to derive the equally valuable information of memory technologies that would \textit{not} be applicable. }

{We provide two illustrative examples derived from \cref{fig:transformer-mem-bw-scaling}:

\noindent\textit{Ex. 1:} The theoretically optimal MP8\_DP128 configuration is achievable with a hybrid memory system offering at least $4.25\times$ higher aggregate capacity than the baseline (cf. \cref{fig:baseline-TT-memFP-breakdown}), and outperforms the baseline if the expanded memory is accessible at a bandwidth of at least 500GB/s.

\noindent\textit{Ex. 2:} A system architect considering CXL-attached memory can quickly deduce that, in order to benefit the given workload, the technology must be capable of delivering 500GB/s to 340GB of memory, at a minimum. In practice, that would require a memory device accessible over 32 lanes of CXL 3.0.}

\medskip
\subsubsection{Effect of Per-node Compute Capability}
\label{sec:transformer-compute}

The tremendous demand for DL training drives rapid evolution of the hardware used in clusters deployed for such purpose.
\TheName can be used to study the effect of per-node compute capability by scaling each node's peak performance (${perf}_{max}$) to model hardware of different generations.

\begin{figure}[t]
    \centering    
    \includegraphics[width=0.95\columnwidth]{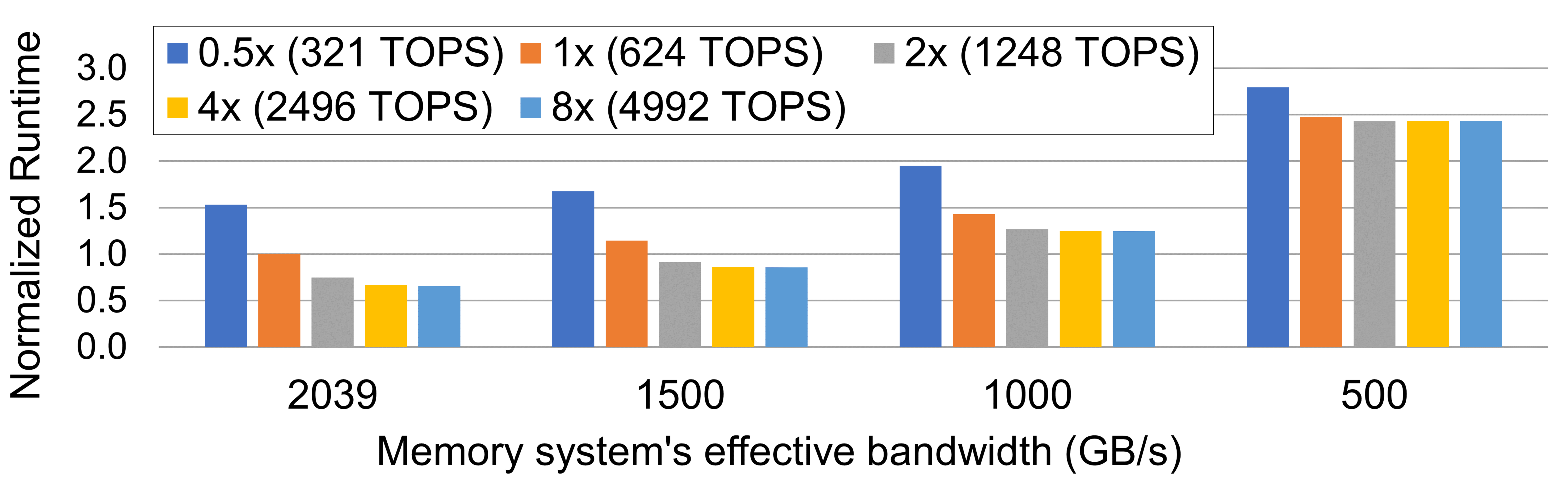}
    \vspace{-2mm}
    \caption{Effect of node compute capability relative to baseline A100 GPU (MP8\_DP128 configuration).}
    \label{fig:baseline-compute-scaling}
    \vspace{-3mm}
\end{figure}

\cref{fig:baseline-compute-scaling} shows the effect of per-node compute scaling on the MP8\_DP128 configuration, assuming a memory system of varying $bw_{hybrid}$ as a function of $bw_{EM}$, and sufficient capacity to hold the model.
At the highest $bw_{EM}$ of 2TB/s, halving compute capability---e.g., by replacing the baseline A100 with a lower-end GPU---increases the runtime by $50\%$, while doubling it reduces the runtime by 25\%. 
Scaling compute further has diminishing returns, as training becomes communication bound.
For lower memory bandwidth availability, the impact of compute capability scaling further diminishes due to the additional bottleneck of memory bandwidth. %
{\textit{System designers can employ such  studies to predict the impact of a next-generation GPU on a cluster's overall performance.}}

\medskip
\subsubsection{Effect of Networking Capability}
\label{sec:transformer-networking}

A cluster's network bandwidth plays a critical role in the overall training time.
Especially at higher MP configurations, training time is dominated by the exposed blocking communication patterns in forward and input gradient phases.
In our modeled cluster, there is an \textit{intra-pod} bandwidth to communicate with GPUs within a pod and a lower \textit{inter-pod} bandwidth for communication across pods (see \cref{tab:baseline-param}).
We use a \textit{Hierarchical Collective} implementation \cite{cho:blueconnect, rashidi:themis} in our evaluations, which first reduces data across GPUs of the same pod, followed by inter-pod reduction. %
Such local network bandwidth-aware collective optimization reduces the communication volume on the lower-bandwidth inter-pod links. %

\begin{figure}
    \vspace{-2mm}
    \centering
        \subfloat[MP64\textunderscore DP16] {
    	    \includegraphics[width=0.49\columnwidth]{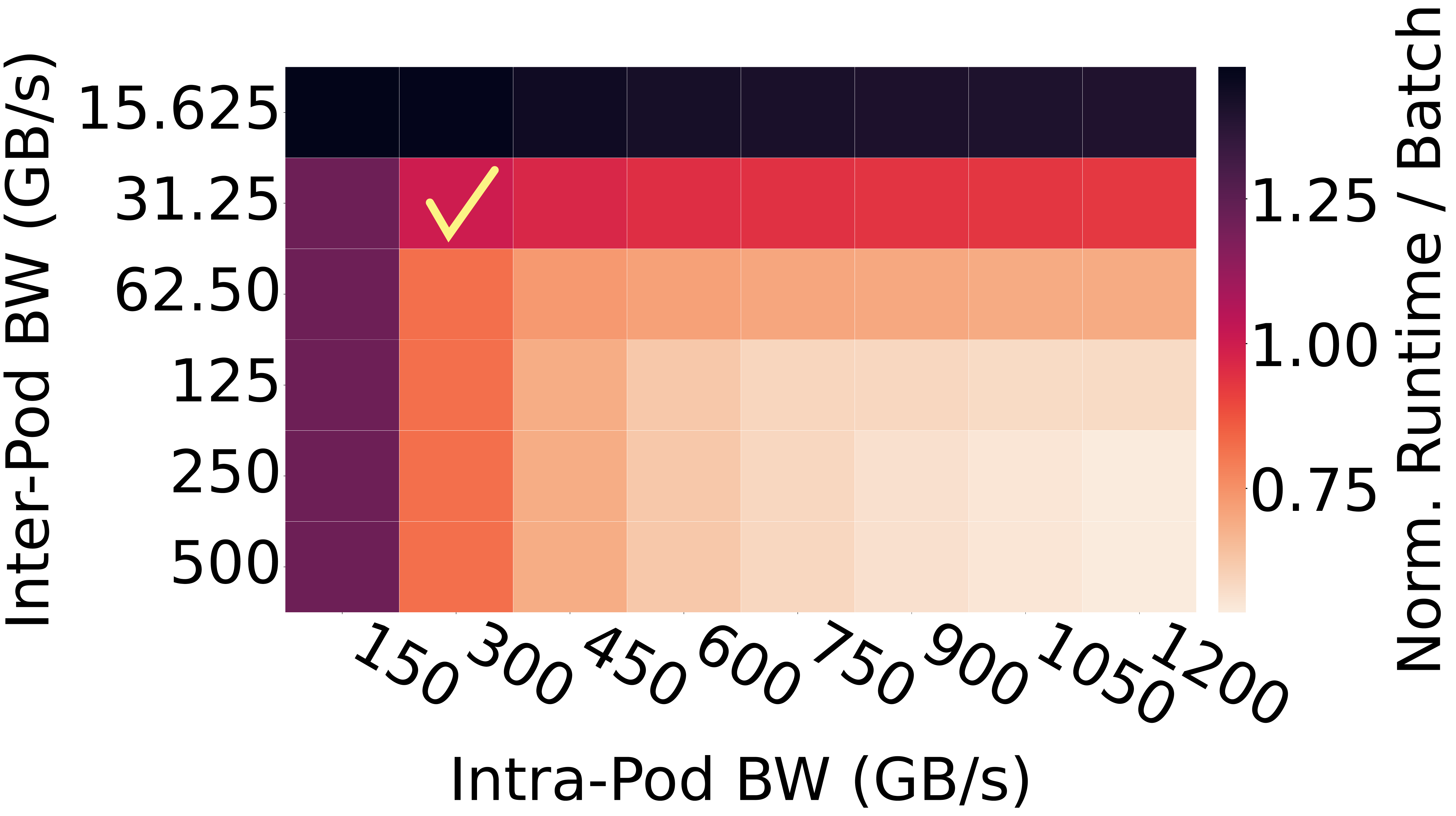}
    	    \label{fig:MP64_DP16_nw_scaling}
    	}
        \subfloat[MP8\textunderscore DP128] {
    	    \includegraphics[width=0.49\columnwidth]{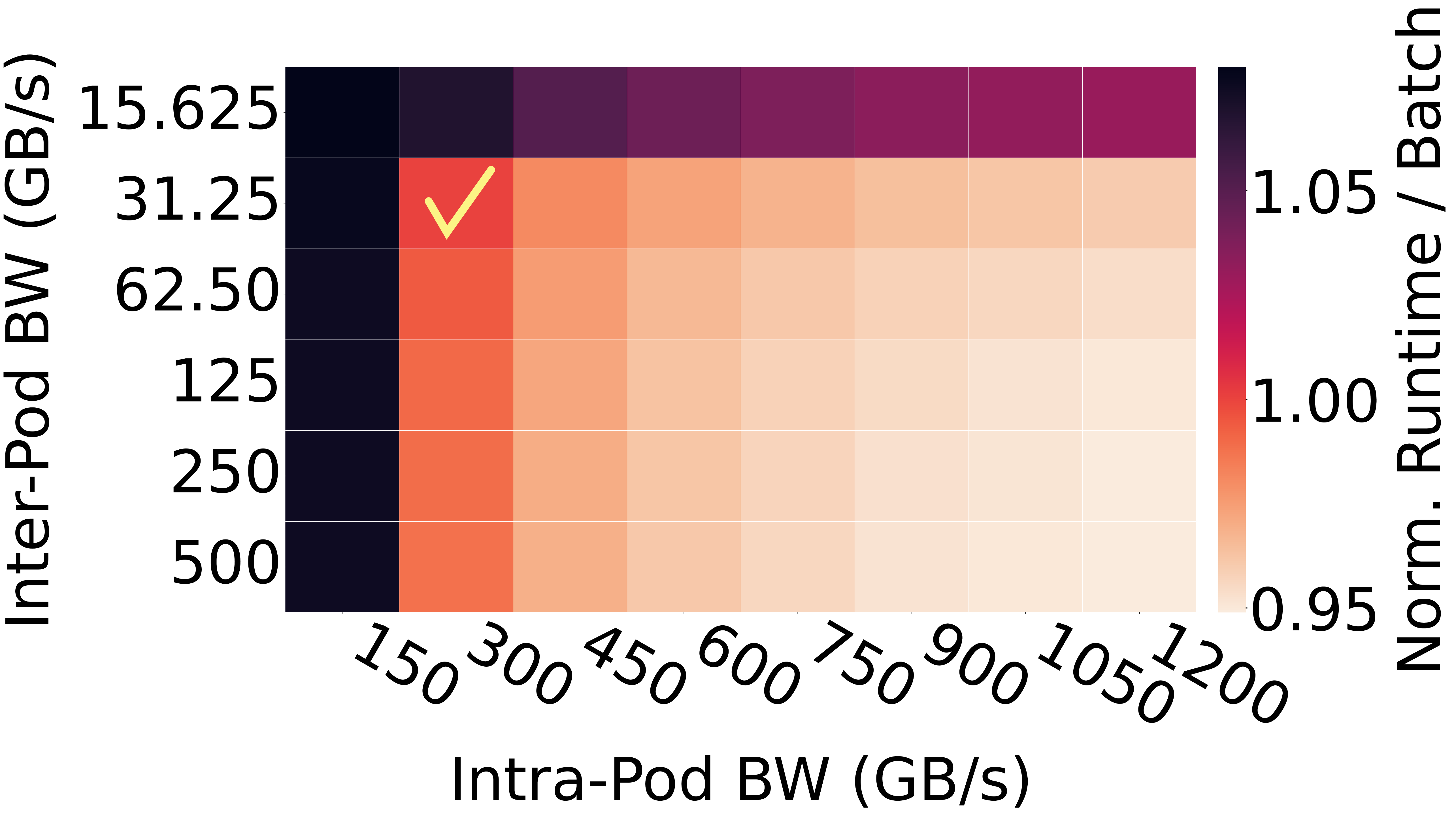}
    		\label{fig:MP8_DP128_nw_scaling}
    	}
    \caption{Effect of network capabilities on cluster-scale performance. The check-mark indicates the baseline configuration to which all the other values are normalized. {300/31.25 GB/s is currently a typical NVLink/InfiniBand configuration.}
    }
    \label{fig:transformer-network-scaling}
    \vspace{-3mm}
\end{figure}

\cref{fig:transformer-network-scaling} shows the effect of both intra- and inter-pod network bandwidth scaling on training time for two different (MP, DP) configurations: the communication-bound MP64\_DP16 and the compute-bound MP8\_DP128.
In each plot, we vary the two network bandwidths while keeping the compute and memory bandwidth constant to the baseline system's values (\cref{tab:baseline-param}).

Unsurprisingly, MP64\_DP16 is majorly affected by network bandwidth, \revision{as the MP dimension straddles several 8-node pods, which feature high intra-pod bandwidth}.  
Halving intra-/inter-pod bandwidth results in a 48\%/22\% slowdown, respectively.
On the contrary, doubling intra-/inter-pod bandwidth reduces runtime by 3\%/18\%, respectively, while doubling both reduces runtime by 27\%.
Evidently, the most effective network scaling scales bandwidth on both dimensions, while scaling only one dimension's bandwidth provides reduced or marginal gains.

\revision{This effect is attributed to the mapping of the workload on the underlying cluster. 
The performance-critical all-reduce communication in forward and backward propagation is bottlenecked by \textit{both intra- and inter-pod links}. Thus, increasing only one dimension's capability yields limited gains (as the other dimension remains a bottleneck), while boosting the capabilities of both dimensions has an amplificatory effect.}
In contrast, when MP8\_DP128 training is feasible, the network's role is much less critical. 
Reducing both intra- and inter-pod bandwidth to 50\% only degrades performance by 11\%, while even increasing both by $4\times$ only boosts performance by 5\%.

\begin{figure}
    \centering
    \includegraphics[width=\columnwidth]{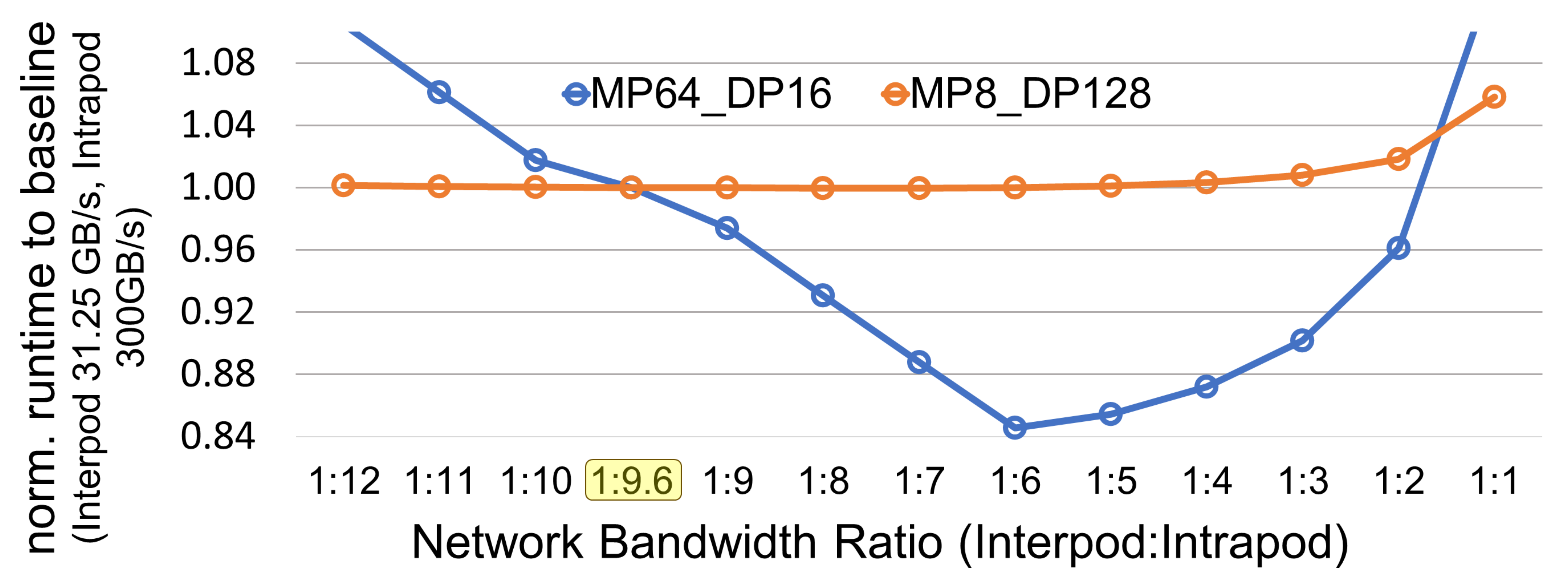}
    \vspace{-4mm}
    \caption{\revision{Impact of relative inter-/intra-pod network bandwidth allocation on training runtime. The total bandwidth is kept constant at 331.25 GB/s across all the ratios. The highlighted 1:9.6 ratio represents the baseline cluster's configuration. }
    }    
    \label{fig:nw-bw-ratio}
\end{figure}

\revision{We conduct an additional experiment to focus on identifying the ideal balance of the bandwidth available at the two dimensions. Instead of varying the two values independently, as previously done in \cref{fig:transformer-network-scaling}, \cref{fig:nw-bw-ratio} shows the performance trends when the available network bandwidth is re-distributed between intra-/inter-pod links, while its aggregate value always remains fixed.
For the baseline MP64\textunderscore DP16 configuration with an aggregate full duplex bandwidth of 331.25 GB/s per direction (300 GB/s intra-pod and 31.25GB/s inter-pod), we find an optimal bandwidth ratio of 1:6 between inter-/intra-pod network bandwidth (i.e., 284 GB/s intra-pod and 47.32 GB/s inter-pod). 
For lower/higher ratios, inter-/intra-pod network traffic becomes a bottleneck, respectively.
In the case of compute-bound MP8\textunderscore DP128, where the MP dimension is entirely contained within a pod, the performance-critical MP communication is entirely dictated by intra-pod bandwidth while the DP communication across the pods is affected by the inter-pod link bandwidth.
Therefore, MP communication is not a bottleneck and performance is largely insensitive to the rebalanced bandwidth ratio.
Performance starts dropping beyond 1:5 (i.e., with intra-pod bandwidth less than 276GB/s), as MP communication starts reappearing as a bottleneck.
Overall, 1:6 inter-/intra-pod network bandwidth provisioning appears to be the ratio that best accommodates both training configurations, improving training time by up to $15\%$ compared to the default 1:9.6 ratio.
}

{\textit{System designers can use such analysis to determine the topology and interconnect technology to use (e.g., Ethernet, InfiniBand, NVLink), while balancing the resulting performance with the associated cost of the selected hardware resources.}}

\subsection{DLRM Evaluation}
\label{sec:dlrm-eval}

\begin{figure}[t]
    \centering 

    \subfloat[Breakdown of comp./comm. time \& per-node memory footprint. ]{
    \includegraphics[width=\columnwidth]{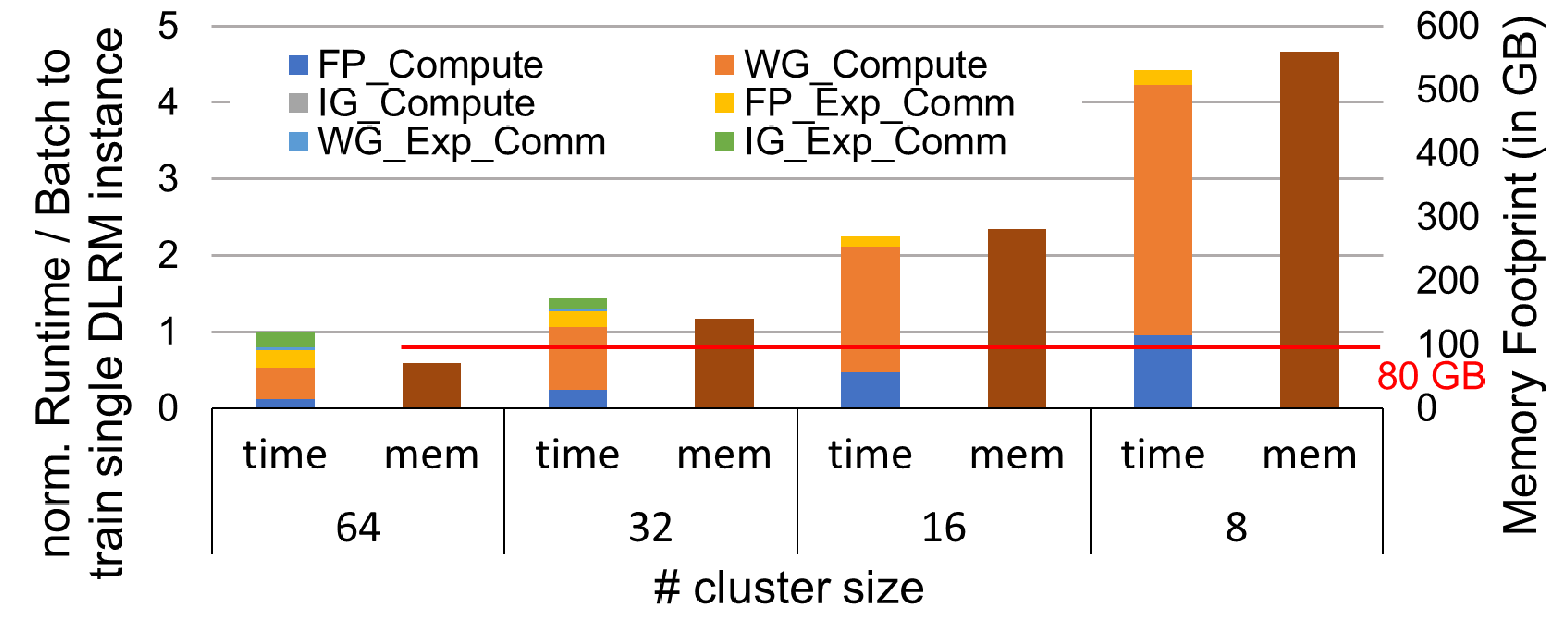}
    \label{fig:DLRM-A_baseline-TT-memFP-breakdown}
    }
\vspace{-2mm}
    
    \subfloat[Effect of memory bandwidth availability to the extended memory system.]{
    \includegraphics[width=\columnwidth]{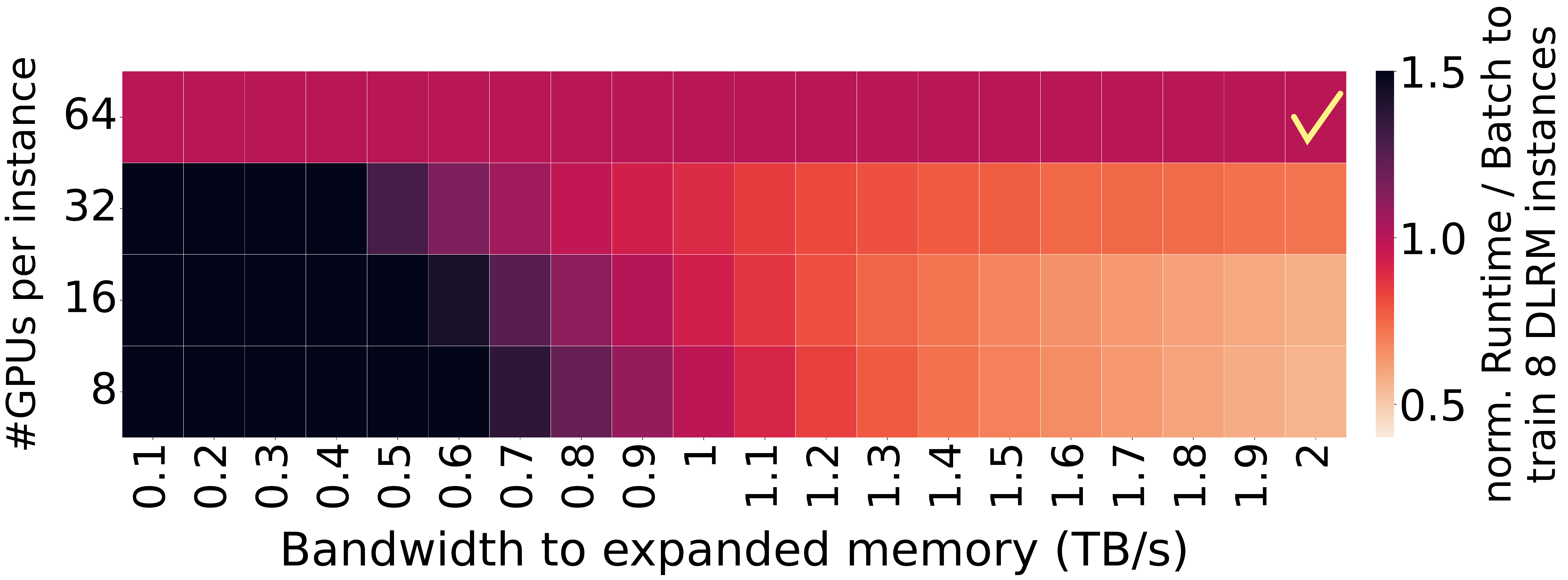}    
    \label{fig:DLRM-A_mem_bw_analysis}
    }
    \caption{DLRM's training performance normalized to 64 nodes with 2TB/s memory bandwidth.     }
    \vspace{-4mm}
\end{figure}

\begin{table*}[b]
    \centering
    \begin{footnotesize}
    \begin{center}
    \caption{Per-node details of various cluster configurations. }
    \vspace{-2mm}
    \label{tab:dgx-variants}
    \scalebox{0.95}{
        \begin{tabular}{|c|c|c|c|c|c|c|p{5cm}|}
        \hline
        \multirow{2}{*}{\textbf{config}} & 
        \multicolumn{2}{c|}{\textbf{compute}} & 
        \multicolumn{4}{c|}{\textbf{
memory}} & 
        \multirow{2}{*}{\textbf{network (topology : bandwidth per node)}} \\
        \cline{2-7} & 
        \textbf{node} & 
        \textbf{Peak TOPS} &
        \textbf{local cap. (GB)} &
        \textbf{local bw (GB/s)} &
        \textbf{exp. cap. (GB)}&
        \textbf{exp. bw (GB/s)}&\\
        \hline  
            A0 & \multirow{3}{*}{V100} & \multirow{3}{*}{125} & \multirow{3}{*}{80$^*$} & \multirow{3}{*}{900} & 0 & 0 & two-level switch :  \\
            \cline{1-1}\cline{6-7}
            A1 &  &  &  &  & 480 & 500 & \multirow{2}{*}{150 GB/s intra-pod, 6.25 GB/s inter-pod} \\
            \cline{1-1}\cline{6-7}
            A2 &  &  &  &  & 201 & 1000 &  \\
            \hline\hline
            B0 & \multirow{3}{*}{A100} & \multirow{3}{*}{625} & \multirow{3}{*}{80} & \multirow{3}{*}{2039} & 0 & 0 & two-level switch :  \\
             \cline{1-1}\cline{6-7}
            B1 &  &  &  &  & 480 & 500 & \multirow{2}{*}{300 GB/s intra-pod, 31.25 GB/s inter-pod} \\
             \cline{1-1}\cline{6-7}
            B2 &  &  &  &  & 201 & 1000 &  \\
            \hline\hline
            C0 & \multirow{3}{*}{H100} & \multirow{3}{*}{1979} & \multirow{3}{*}{80} & \multirow{3}{*}{3350} & 0 & 0 & two-level switch :  \\
            \cline{1-1}\cline{6-7}
            C1 &  &  &  &  & 480 & 500 & \multirow{2}{*}{450 GB/s intra-pod, 62.5 GB/s inter-pod} \\
            \cline{1-1}\cline{6-7}
            C2 &  &  &  &  & 201 & 1000 &  \\
            \hline\hline
            Dojo & Tray & 54,300 & 640 & 16000 & 0 & 0 & one-level switch : $20\times50GB/s$ per direction \\
            \hline\hline
            TPU v4 & TPU & 275 & 32 & 1200 & 39 & 1200 & 3D torus : $6\times48 GB/s$ per direction \\
            \hline
            \multicolumn{8}{l}{$^*$Although the V100 GPU features 32GB of memory, we model 80GB instead to keep the memory system configuration options of clusters A, B, and C aligned.}\\
    \end{tabular}    
    }
    \vspace{-5mm}
    \end{center}
    \end{footnotesize}

\end{table*}

We now briefly cover evaluation highlights for the training of a 1.2 trillion parameter DLRM, modeled as described in Table V of Rashidi et al. \cite{rashidi:2020}.
Fig. \ref{fig:DLRM-A_baseline-TT-memFP-breakdown} shows the training time breakdown for single DLRM instance and corresponding per-node memory footprint for different cluster sizes.
Since the DLRM's memory footprint is relatively smaller than \cref{sec:transformer-eval}'s Transformer-1T model, we start with a smaller cluster comprising only 8 pods of our baseline DGX cluster (64 GPUs in total).
As the cluster's size decreases,  %
the exposed communication delay decreases at the cost of increased memory footprint per node required, resulting in a compute delay increase.
However, the overall increase in training time is sublinear with the node count reduction, especially in the 64--16 range.
Thus, \textit{memory expansion can not only be used to improve DLRM training efficiency, but also better performance for a training workload comprising several DLRM models.} This is a common use case, as large corporations often need to train multiple DLRMs for different purposes \cite{acun:2021, mudigere:2022}.

\cref{fig:DLRM-A_mem_bw_analysis} evaluates the overall turnaround time of training 8 DLRMs on 64 GPU nodes as a function of available bandwidth to the expanded memory. 
\textit{While DLRM's performance is more sensitive to memory bandwidth, results qualitatively match \cref{sec:transformer-memory}'s takeaways. }
Performance improvement opportunities require memory expansion solutions delivering at least {75\%} additional memory capacity at 800GB/s; a 200GB expanded memory accessible at {1.5TB/s} improves training time by {1.5$\times$}.

\subsection{Comparative DL Training on Different Clusters}
\label{sec:cluster-comparison}

\begin{figure}
    \centering    
    \subfloat[Cluster of 1024 A100 GPUs, grouped in 128 8-GPU pods.]{
    \includegraphics[width=.8\columnwidth]{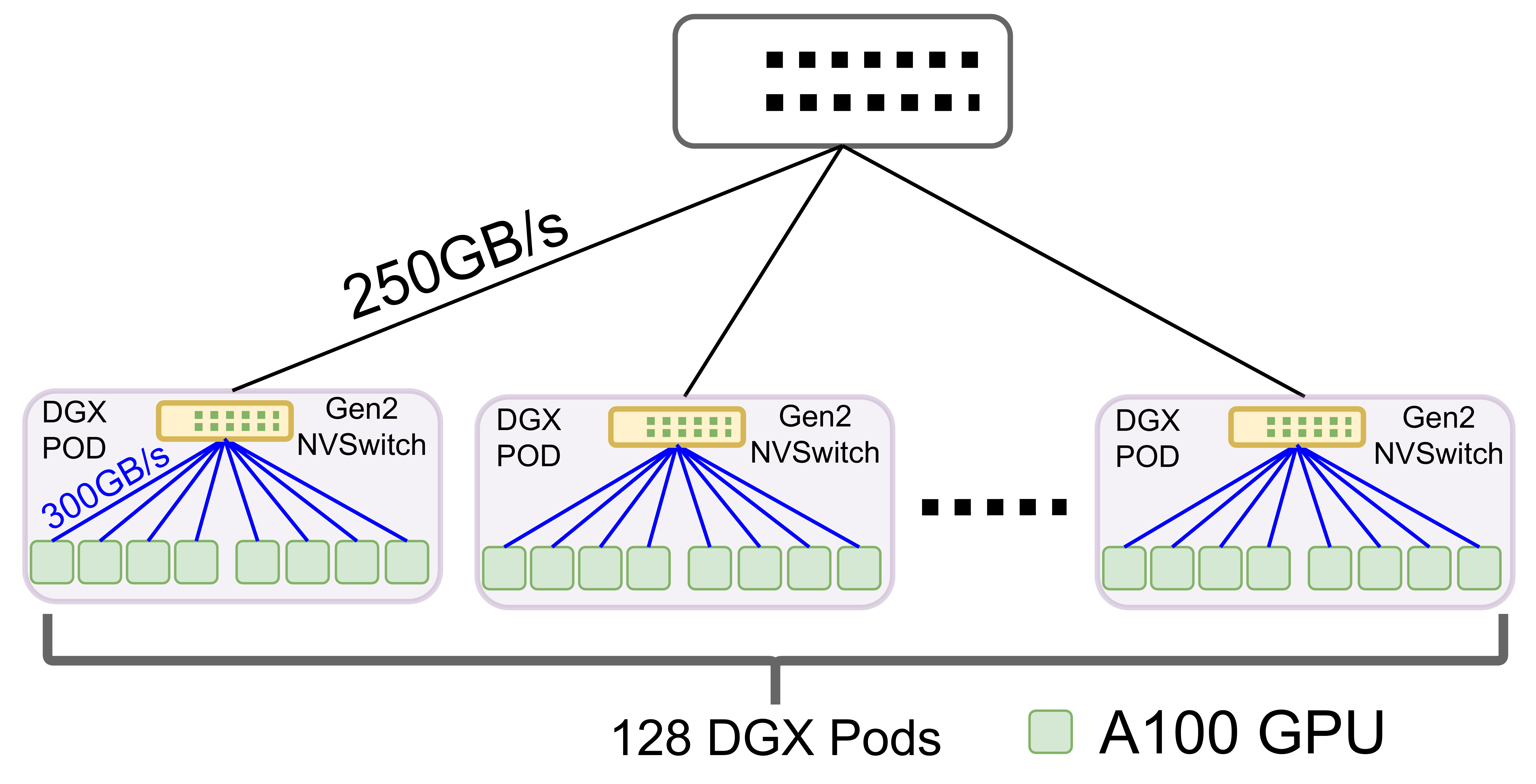}    
    \label{fig:apx:DGX}
    }
    
    \subfloat[Cluster of 4096 TPU v4 chips connected in 3D Torus topology. ]{
    \includegraphics[width=.7\columnwidth]{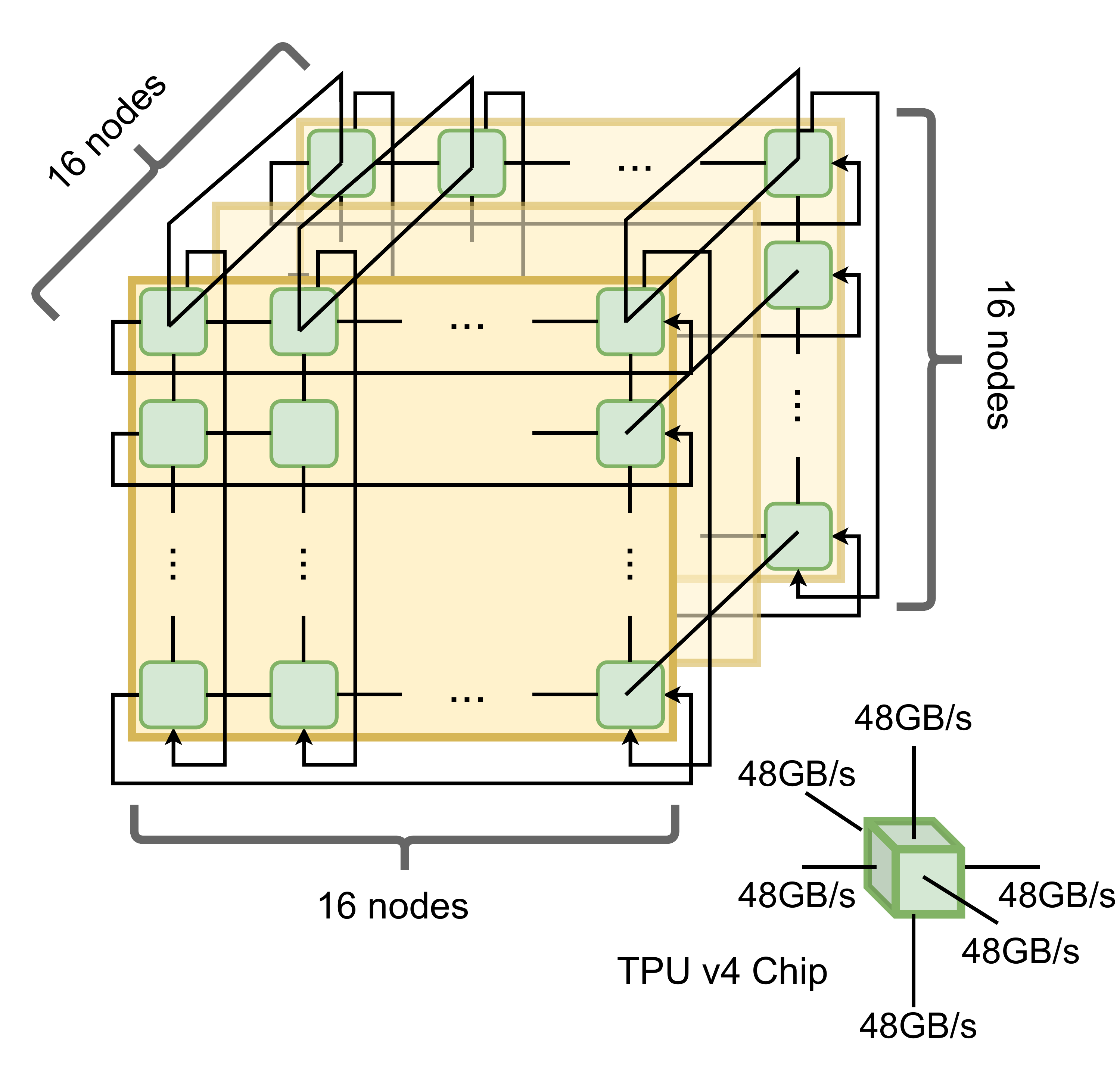}   
    \label{fig:apx:TPU-cluster}
    }
    
    \subfloat[Cluster of 64 Dojo D1 Training Matrices connected by a switch. ]{
    \includegraphics[width=.5\columnwidth]{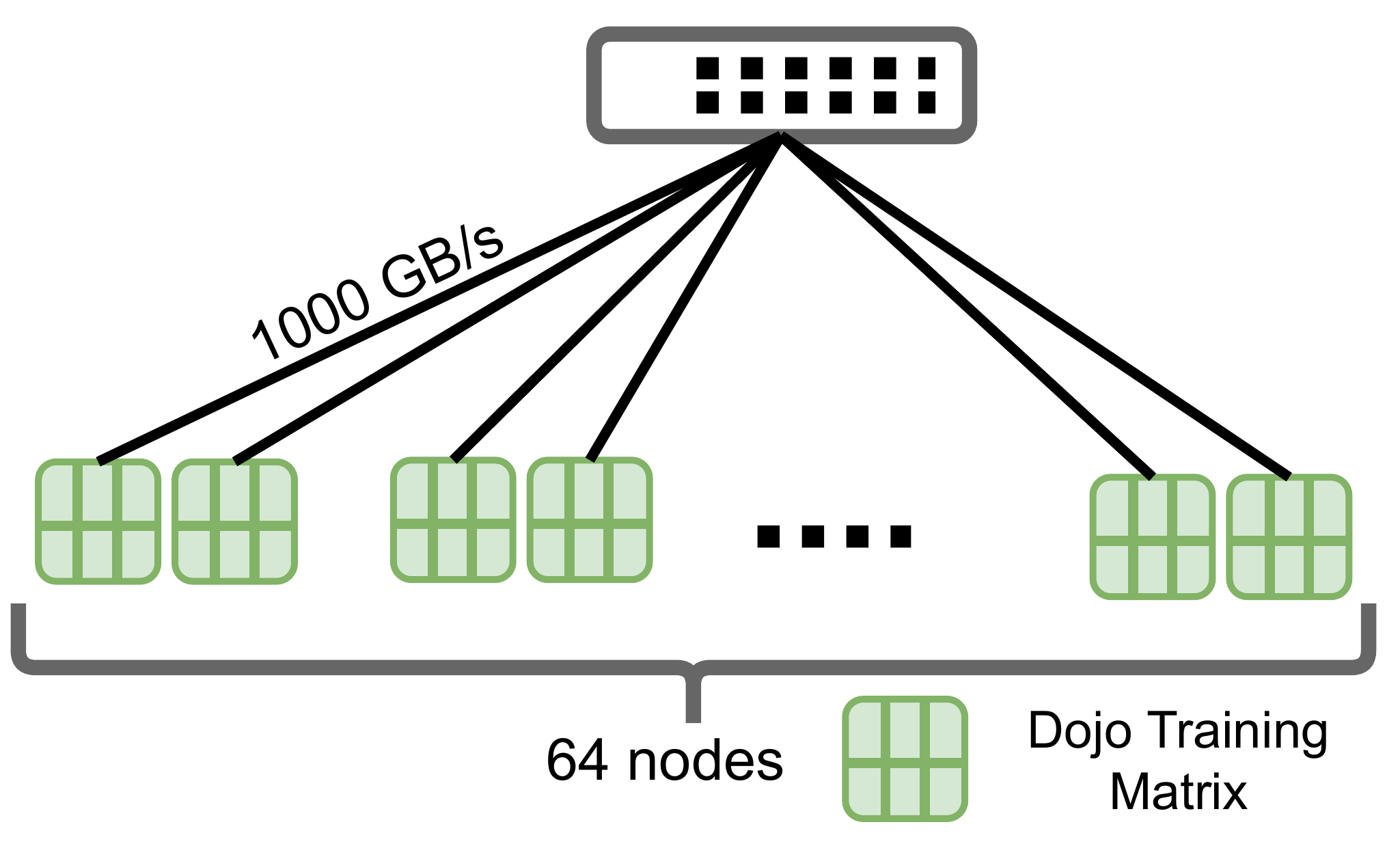}    
    \label{fig:apx:dojo-cluster}
    }
    \caption{Illustration of three cluster types evaluated in \cref{sec:cluster-comparison}. All link bandwidths shown are per direction.   }
    \label{fig:all-clusters}
\end{figure}

We conclude \TheName's utility demonstration by comparing 11 different DL training clusters, summarized in \cref{tab:dgx-variants}: nine GPU-based clusters, a Google TPU v4, and a Tesla Dojo cluster. 
The general structure of the three cluster types is depicted in \cref{fig:all-clusters}.
Our goal is \textit{not} to declare the ``best'' system---as they drastically differ in cost, node count, definition of a ``node'', etc.---but to demonstrate the different behavior across three very dissimilar clusters when training the same huge model.

\begin{figure*}[h]
    \centering
    \includegraphics[width=\textwidth]{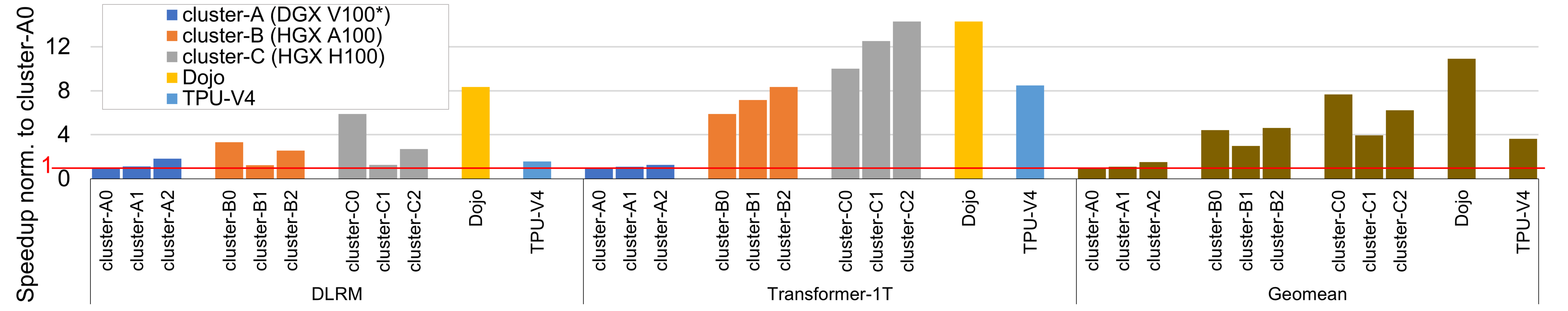}
     \vspace{-6mm}
    \caption{Comparison of runtime across different cluster configurations, normalized to cluster A0.}
    \label{fig:nvidia-cluster-comp}
     \vspace{-4mm}
\end{figure*}

\paragraph{DGX cluster variants} 
We model three base 1024-GPU cluster variants---A, B, and C---with different resource (compute, memory, network) provisioning. We base each design on a major GPU model (V100, A100 and H100).
All GPU cluster variants are organized in 16-GPU pods and feature a two-dimensional network like the one shown in \cref{fig:DGX}.
For each base cluster variant, we evaluate three memory systems---0, 1 and 2---with different characteristics, for a total of nine GPU cluster variants.
Memory system 0 consists of only local GPU memory without any capacity expansion.
Memory systems 1 and 2 are hypothetical expanded memory systems accessible at 0.5TB/s and 1TB/s, respectively.

\paragraph{TPU v4 cluster}
We model a TPU cluster of 4096 TPU-V4 chips connected in a 3D Torus topology with 48GB/s full duplex links. %
Each TPU features 32MB of on-chip SRAM, a 32GB HBM with a memory bandwidth of 1.2TB/s, and 275 TFLOPS peak performance \cite{tpuv4}.

\paragraph{Dojo cluster}
We model a Dojo cluster of 64 nodes (``trays'', each comprising several training tiles and interface processors). 
Each node has 66GB of on-chip SRAM, 640GB of memory with a 16 TB/s bandwidth, and 54.3 PFLOPS peak performance.
Given limited publicly available information, we model the  network topology as a single logical switch delivering 1 TB/s of full duplex network bandwidth to each node \cite{lambert:2022}.

\smallskip
\cref{fig:nvidia-cluster-comp} shows the speedup for DLRM and Transformer-1T across different cluster configurations. %
Every cluster except A0, B0, C0, and Dojo assumes per-node memory expansion with capacity and bandwidth characteristics listed in \cref{tab:dgx-variants}.
DLRM training is modeled as described in \cref{sec:dlrm-eval}:
Clusters A0, B0, and C0 use 64 nodes to run a single DLRM instance.
A1, B1 and C1 leverage their expanded memory to train one DLRM per 16 nodes.
Likewise, A2, B2 and C2 use 8 nodes per instance.
The reported speedup for DLRM refers to training a total of 8 model instances. 
For Transformer-1T, speedup refers to training a single instance on the entire cluster. All reported speedups are normalized to cluster A0 as baseline.

Clusters A2, B2 and C2 benefit from their expanded memory at 1TB/s, delivering $1.8\times$, $2.6\times$, and $2.7\times$ speedups, respectively, for DLRM. 
While clusters A1, B1 and C1 fare poorly for DLRM due to their lower bandwidth to expanded memory, B1 and C1 deliver a good speedup of $7.2\times$ and $12.5\times$, respectively, for Transformer-1T. 
Increasing expanded memory bandwidth to 1TB/s further improves speedup for  Transformer-1T (e.g., to $14.3\times$ for C2).

Cluster A2's double bandwidth to expanded memory improves cluster A1's DLRM performance by $1.64\times$.
Due to the memory-bound nature of DLRM, memory bandwidth improvements are more critical than compute capacity or network bandwidth.
On the other hand, Transformer-1T is more sensitive to the compute capacity and network bandwidth, and therefore low compute and network bandwidth drastically reduces speedup opportunities for clusters A1 and A2.
Transformer-1T benefits from TPU's large compute capacity and network bandwidth, but the DLRM suffers from its low memory capacity and local memory bandwidth. In contrast, Dojo significantly benefits both workloads due to large on-chip SRAM, memory capacity, and high network bandwidth. 

Overall, among GPU cluster variants, there is no single optimal configuration for both Transformer-1T and DLRM, as different workloads are impacted differently. Disregarding any cost considerations, the best GPU cluster on average is C0, delivering a 7.7$\times$ speedup over the baseline A0 cluster. \textit{Memory expansion is an effective technique for all clusters when training Transformers, but only for the lowest-end cluster A on average, due to DLRM's memory bandwidth sensitivity.} %
\revision{This study demonstrates that, as one size \textit{does not} fit all, system architects should evaluate a workload mix representative of the main use cases for the target cluster to determine the system configuration that best fits the ensemble.}

\subsection{\TheName takeaways: versatility and speed}
\label{sec:versatility}

{
The extensive evaluation in \cref{sec:transformer-eval,sec:dlrm-eval} demonstrates \TheName's versatility and the breadth of case studies it facilitates. 
We illustrated that \TheName enables joint sensitivity analysis of the effect of node compute capability, memory system design, and network provisioning on cluster performance, facilitating balanced resource provisioning and identification of cost-reduction opportunities. 
\TheName also allows rapid exploration and evaluation of memory system design on a cluster's performance as a function of its capacity and bandwidth characteristics, helping system designers determine what existing technologies for memory expansion are viable to improve training performance, and gauge the impact of relevant future technologies.

While \cref{sec:case-studies}'s evaluation focused on demonstrating the utility and flexibility of the tool, an additional key strength of \TheName is the short turnaround time of experiments, allowing researchers to rapidly glean performance trends. }
Once a target model is broken down 
into its layer-wise representation as specified in \cref{sec:method:workload-modeling}, exploring a broad design space is a matter of few hours.
\TheName's methodology allows modeling different compute nodes, network topologies, memory bandwidths and parallelization strategies in an embarrassingly parallel fashion on commodity processors.
To provide some concrete data points, generating the two memory bandwidth sensitivity heatmaps (\cref{fig:transformer-mem-bw-scaling} for Transformer-1T and \cref{fig:DLRM-A_mem_bw_analysis} for DLRM) takes about 5 hours / 45 minutes, respectively, on a single 24-core Intel Xeon Silver server.
Other analyses presented in the paper require comparable runtimes.
Such rapid exploration of a wide range of design choices is a valuable capability \TheName contributes.
\section{Conclusion}
\label{sec:conclusion}

We introduced \TheName, a holistic, end-to-end methodology and workflow for rapid design space exploration of key parameters affecting the performance and efficiency of large-scale distributed DL training: the model's parallelization strategy and provisioning of each of the cluster's key resources.
We demonstrated \TheName's utility with case studies on DLRM and Transformer models, deriving actionable cluster design hints for system designers, and identifying required bandwidth and capacity characteristics for emerging memory expansion techniques to have a positive impact on cluster performance on distributed DL training.
\TheName enables the evaluation of a wide spectrum of key cluster design parameters in a matter of few hours using very modest physical hardware resources.

\section*{Acknowledgements}
\label{sec:acks}
We thank William Won, Geonhwa Jeong, Shreerang Dabade and Raveesh Garg for their help in the early stages of constructing the paper's methodology, as well as Marina Vemmou and Hamed Seyedroudbari for their feedback on drafts of the paper.
We also thank Sudarshan Srinivasan for technical discussions on this work. 
This work was generously supported by a research gift from Samsung Semiconductor's Memory Solutions Lab.

\bibliographystyle{IEEEtranS}
\balance
\bibliography{COMET/main}

\end{document}